\documentclass[useAMS]{mn2e}

\usepackage{latexsym,graphicx}

\usepackage{color}

\newcommand\kms{{\rm\,km\,s^{-1}}}
\newcommand\msun{\rm\,M_\odot}
\newcommand\lsun{\rm\,L_\odot}
\newcommand\rsun{\rm\,R_\odot}
\newcommand\myr{\msun \, {\rm yr}^{-1}}
\newcommand{\MC}{\multicolumn}
\newcommand\hii{H\,{\sc ii} \,}

\def\apgt{\ {\raise-.5ex\hbox{$\buildrel>\over\sim$}}\ }
\def\aplt{\ {\raise-.5ex\hbox{$\buildrel<\over\sim$}}\ }

\title[Spectroscopy of Sk$-$69$\degr$\,279 and its shell]{Optical spectroscopy of the blue supergiant 
Sk$-$69$\degr$\,279 and its circumstellar shell with SALT}\author[V. V.~Gvaramadze et al.]
        {V. V.~Gvaramadze,$^{1,2,3}$\thanks{E-mail: vgvaram@mx.iki.rssi.ru} A. Y.~Kniazev,$^{1,4,5}$ O. V.~Maryeva,$^{6,7}$ 
        \newauthor and L. N.~Berdnikov$^{1,3,8}$\\
        $^{1}$Sternberg Astronomical Institute, Lomonosov Moscow State University, Universitetskij Pr. 13, Moscow 119992, Russia\\
        $^{2}$Space Research Institute, Russian Academy of Sciences, Profsoyuznaya 84/32, 117997 Moscow, Russia \\
        $^{3}$Isaac Newton Institute of Chile, Moscow Branch, Universitetskij Pr. 13, Moscow 119992, Russia \\
        $^{4}$South African Astronomical Observatory, PO Box 9, 7935 Observatory, Cape Town, South Africa \\
        $^{5}$Southern African Large Telescope Foundation, PO Box 9, 7935 Observatory, Cape Town, South Africa \\
        $^{6}$Astronomical Institute, Czech Academy of Sciences, Fri\v{c}ova 298, 251 65 Ond\v{r}ejov, Czech Republic \\ 
        $^{7}$Special Astrophysical Observatory of the Russian Academy of Sciences, Nizhnii Arkhyz, 369167, Russia\\
        $^{8}$Astronomy and Astrophysics Research Division, Entoto Observatory and Research Center, PO Box 8412, Addis Ababa, Ethiopia\\        
        }
\begin{document}

\date{Accepted 2017 October 31. Received 2017 October 31; in original form 2017 September 23}

\maketitle

\label{firstpage}

\begin{abstract}
We report the results of optical spectroscopy of the blue supergiant Sk$-$69$\degr$\,279 and its 
circular shell in the Large Magellanic Cloud (LMC) with the Southern African Large Telescope (SALT). 
We classify Sk$-$69$\degr$\,279 as an O9.2\,Iaf star and analyse its spectrum by using the 
stellar atmosphere code {\sc cmfgen}, obtaining a stellar temperature of $\approx$30\,kK, a 
luminosity of $\log(L_*/\lsun)=5.54$, a mass-loss rate of $\log(\dot{M}/\myr$)=$-$5.26, and a wind 
velocity of $800 \, \kms$. We found also that Sk$-$69$\degr$\,279 possesses an extended atmosphere 
with an effective temperature of $\approx$24\,kK and that its surface helium and nitrogen abundances 
are enhanced, respectively, by factors of $\approx$2 and 20--30. This suggests that 
either Sk$-$69$\degr$\,279 was initially a (single) fast-rotating ($\ga400 \, \kms$) star, which only recently 
evolved off the main sequence, or that it is a product of close binary evolution. The long-slit 
spectroscopy of the shell around Sk$-$69$\degr$\,279 revealed that its nitrogen abundance is enhanced by 
the same factor as the stellar atmosphere, which implies that the shell is composed mostly of the CNO
processed material lost by the star. Our findings support previous propositions that some massive 
stars can produce compact circumstellar shells and, presumably, appear as luminous blue variables while they are 
still on the main sequence or have only recently left it.
\end{abstract}


\section{Introduction}
\label{sec:int}

Mass loss from massive stars results (under proper conditions) in the formation of compact (pc-scale) circumstellar 
nebulae of various morphologies. Detection of CNO processed material in some of these nebulae by means of 
spectroscopic observations might indicate that their associated stars are at advanced stages of evolution, 
although it is also possible that massive stars can eject nebulae while they are relatively unevolved, i.e. 
still on the main sequence (e.g. Lamers et al. 2001). Among the massive stars, the compact circumstellar 
nebulae are most often found around luminous blue variables (LBVs). Namely, observations show that about 60 
per cent of known bona fide and candidate LBVs are surrounded by such nebulae (Nota et al. 1995; Clark, Larionov 
\& Arkharov 2005). This is contrasted by the detection of only a few circumstellar nebulae around Wolf-Rayet (e.g. 
Marston 1995) and other types of evolved massive stars, and suggests that the duration of the LBV phase is 
comparable to the lifetime of their circumstellar nebulae ($\sim10^4$ yr) and is much shorter than that of other 
major evolutionary phases in the life of massive stars. 

A detection of compact nebulae can be considered as an indication that their associated stars are massive and 
(presumably) evolved (e.g. Clark et al. 2003). Indeed, untargeted searches for compact infrared nebulae in 
the archival data of the {\it Spitzer Space Telescope} and the {\it Wide-field Infrared Survey Explorer} (e.g. 
Gvaramadze, Kniazev \& Fabrika 2010; Wachter et al. 2010; Gvaramadze et al. 2012a) led to the discovery of 
many dozens of such stars, of which, as expected, the most numerous are LBV-like stars (Kniazev \& Gvaramadze 
2015; Gvaramadze \& Kniazev 2017, and references therein). With newly detected candidate and bona fide LBVs, 
the percentage of these stars with circumstellar nebulae has increased to more than 70 per cent (Kniazev, 
Gvaramadze \& Berdnikov 2015; Gvaramadze \& Kniazev 2017). 

The presence of compact nebulae around hot, luminous (non-Wolf-Rayet) stars could be used to consider them as 
ex-/dormant LBVs even if they currently do not show significant spectroscopic and photometric variability 
(e.g. Bohannan 1997). On the other hand, it is still not fully clear at what point(s) in the evolution the massive 
stars become LBVs and produce their nebulae. Although it is likely that many LBVs are at advanced stages of 
stellar evolution, there are also observational hints that the LBV phenomenon might be related to stars 
which only recently left the main sequence or even still on it, and that the LBV-like nebulae might be ejected 
{\it before} their underlying stars become LBVs (Lamers et al. 2001). A possible example of such a star -- the 
blue supergiant Sk$-$69$\degr$\,279 in the Large Magellanic Cloud (LMC) -- is the subject of this paper.

In Section\,\ref{sec:sk}, we present Sk$-$69$\degr$\,279 and its circumstellar shell. Section\,\ref{sec:obs} 
describes our spectroscopic and photometric observations. The spectral classification of Sk$-$69$\degr$\,279 and 
the results of modelling of its spectrum with the stellar atmosphere code {\sc cmfgen} are given, respectively, 
in Sections\,\ref{sec:class} and \ref{sec:mod}. Spectroscopy of the circumstellar shell is discussed in 
Section\,\ref{sec:shell}. In Section\,\ref{sec:dis}, we discuss the obtained results and some issues related to 
the content of the paper. 

\begin{figure*}
\includegraphics[width=15cm]{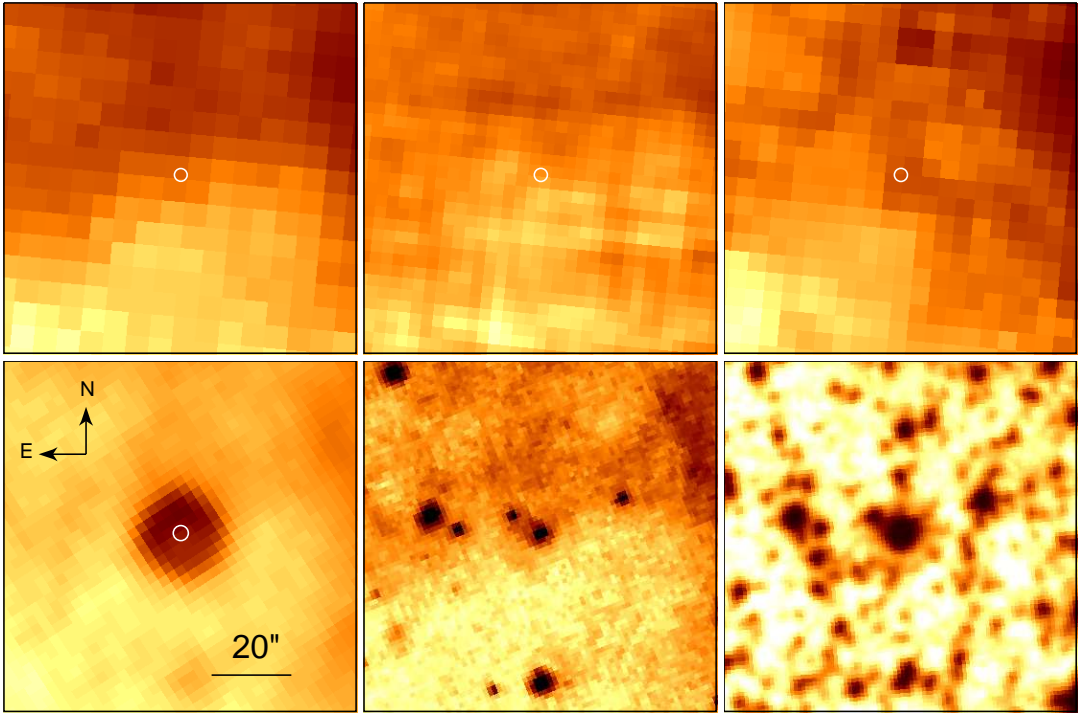}
\centering \caption{From left to right, and from top to bottom: {\it Herschel} 250 and 160 $\mu$m, {\it Spitzer} 
70, 24 and 8\,$\mu$m, and DSS-II red-band images of the region of the LMC containing Sk$-$69$\degr$\,279 (indicated 
by a circle) and its circular shell. The orientation and the scale of the images are the same. At the distance of 
the LMC of 50 kpc, 20 arcsec corresponds to $\approx$4.8 pc.} \label{fig:shell}
\end{figure*}

\section{Sk$-$69$\degr$\,279 and its circumstellar shell}
\label{sec:sk}

Sk$-$69$\degr$\,279 was identified as an OB star by Sanduleak (1970) using an objective prism survey for LMC 
members. Bohannan \& Epps (1974) found in the objective prism spectrum of this star (object \#619 in their survey 
of H$\alpha$ emission-line stars in the LMC) the H$\alpha$ emission line and an unspecified sharp emission feature
bluewards of this line. Later on, Rousseau et al. (1978) classified Sk$-$69$\degr$\,279 as O--B0 (also based on an 
objective prism spectrum), while Conti, Garmany \& Massey (1986) refined the spectral type of this star to O9\,f
using slit spectroscopy. Smith Neubig \& Bruhweiler (1999) classified Sk$-$69$\degr$\,279 as a B0\,II star using 
its {\it International Ultraviolet Explorer} ({\it IUE}) spectrum. This classification was indicated as tentative 
(i.e. B0\,II:) because the luminosity class assigned by the ultraviolet (UV) spectral features does not agree with 
that implied by the derived absolute magnitude. Also, Smith Neubig \& Bruhweiler (1999) found that the B star 
luminosity indicators -- the Si\,{\sc iv}, C\,{\sc iv}, Al\,{\sc iii} and Fe\,{\sc iii} lines -- are weak in the 
spectrum of Sk$-$69$\degr$\,279, which is a characteristic of most confirmed and candidate LBVs in their data set. 
The LBV nature of Sk$-$69$\degr$\,279 is also suggested by the presence of a circular shell of CNO processed material 
around this star (Weis et al. 1997). But since the star does not show significant photometric variability, at least 
during the last decades, it was considered as an ex-/dormant LBV in the comprehensive study of LBVs in the Milky Way 
and Magellanic Clouds by van Genderen (2000; see also Section\,\ref{sec:phot}). 

The circular shell around Sk$-$69$\degr$\,279 was discovered by Weis et al. (1995) using narrow-band imaging in 
H$\alpha$. In the discovery image, it appears as a complete circular shell with a diameter of 18 arcsec, which at a 
distance to the LMC of 50 kpc (Gibson 2000) corresponds to a linear size of $\approx$4.4 pc. \'Echelle spectroscopy 
of the shell revealed a high [N\,{\sc ii}] $\lambda$6584/H$\alpha$ ratio (a factor of 10 higher than in the background 
\hii region), which was interpreted as an indication that the shell is composed of stellar material (Weis et al. 1997). 
It was also found that the shell expands with a velocity of $14 \, \kms$ and that the systemic velocity of the shell 
of $230 \, \kms$ is offset by $20 \, \kms$ with respect to the local medium (Weis et al. 1997). 

The mid-infrared counterpart of the shell around Sk$-$69$\degr$\,279 was discovered during our search for bow shocks 
generated by runaway massive stars in the Magellanic Clouds using the {\it Spitzer Space Telescope} Legacy Survey called 
``Surveying the Agents of a Galaxy's Evolution" (SAGE\footnote{http://sage.stsci.edu/}; Meixner et al. 2006) (for 
motivation and some results of this search, see Gvaramadze, Kroupa \& Pflamm-Altenburg 2010 and Gvaramadze, 
Pflamm-Altenburg \& Kroupa 2011). The survey provides 3.6, 4.5, 5.8 and 8.0\,$\mu$m images obtained with the Infrared 
Array Camera (IRAC; Fazio et al. 2004) and 24, 70 and 160\,$\mu$m images obtained with the Multiband Imaging Photometer 
for {\it Spitzer} (MIPS; Rieke et al. 2004). Besides detection of bow shocks, we discovered in the LMC two circular shells 
of almost the same angular size, one of which turns out to be associated with the already known optical nebula produced by 
Sk$-$69$\degr$\,279 (Weis et al. 1995), while the second one was previously unknown\footnote{Follow-up spectroscopy of 
this shell showed that it is composed of unprocessed material, while spectroscopy of its central stars led to the discovery 
of a WN3b star in a binary system with an O6\,V star and a separate B0\,V star, which hints at the possibility that these 
stars are members of a previously unrecognized star cluster (Gvaramadze et al. 2014b).}.

Fig.\,\ref{fig:shell} shows {\it Herschel} Space Observatory (Pilbratt et al. 2010) 250 and 160\,$\mu$m, {\it Spitzer} MIPS 
70 and 24 $\mu$m and IRAC 8 $\mu$m, and Digitized Sky Survey II (DSS-II; McLean et al. 2000) red band images of 
Sk$-$69$\degr$\,279 (indicated by a circle) and its circular shell. The shell is clearly seen in the 24\,$\mu$m image 
as a circular nebula of the same angular diameter as the optical one (i.e. $\approx18$ arcsec) with the northern edge somewhat 
brighter than the opposite one. The brightness asymmetry could be understood if the more brighter side of the shell impinges on 
a more denser ambient medium, as is evidenced by the presence of diffuse emission in the north direction attached to the shell 
in the {\it Herschel} 160 $\mu$m and {\it Spitzer} 70 and 8\,$\mu$m images, or might be caused by stellar motion in the 
north direction (cf. Section\,\ref{sec:dis}). Fig.\,\ref{fig:shell} also shows that there is a gleam of emission associated 
with the shell in the {\it Spitzer} 8\,$\mu$m and DSS-II images. 

The details of Sk$-$69$\degr$\,279 are summarized in Table\,\ref{tab:det}. The spectral type is based on our spectroscopic 
observations (see Section\,\ref{sec:class}). The $U$ magnitude is from Zaritsky et al (2004). The $BVI_{\rm c}$ magnitudes 
are the mean values derived from our three CCD measurements in 2013--2016 (see Table\,\ref{tab:phot}). The coordinates
and the $JHK_{\rm s}$ photometry are taken from the Two-Micron All Sky Survey (2MASS) All-Sky Catalog of Point Sources 
(Cutri et al. 2003). The IRAC ($[3.6]$--$[8.0]$) photometry is from Bonanos et al. (2009).

\begin{table}
  \caption{Details of Sk$-$69$\degr$\,279.}
  \label{tab:det}
  \begin{center}
 \begin{tabular}{lrr}
    \hline
  Spectral type & O9.2\,Iaf \\
  RA(J2000) & $05^{\rm h} 41^{\rm m} 44\fs67$  \\
  Dec(J2000) & $-$69$\degr 35\arcmin 15\farcs0$ \\
  $U$ (mag) & 12.08$\pm$0.02 \\
  $B$ (mag) & 12.87$\pm$0.01 \\ 
  $V$ (mag) & 12.81$\pm$0.01 \\
  $I_{\rm c}$ (mag) & 12.69$\pm$0.01 \\
  $J$ (mag) & 12.678$\pm$0.029 \\ 
  $H$ (mag) & 12.583$\pm$0.037 \\
  $K_{\rm s}$ (mag) & 12.528$\pm$0.038 \\
  $[3.6]$ (mag) & 12.286$\pm$0.028 \\ 
  $[4.5$] (mag) & 12.129$\pm$0.029 \\
  $[5.8]$ (mag) & 12.082$\pm$0.043 \\
  $[8.0$] (mag) & 11.829$\pm$0.041 \\
  \hline
 \end{tabular}
\end{center}
\end{table}

\begin{figure}
\begin{center}
\includegraphics[width=5.5cm,angle=270,clip=]{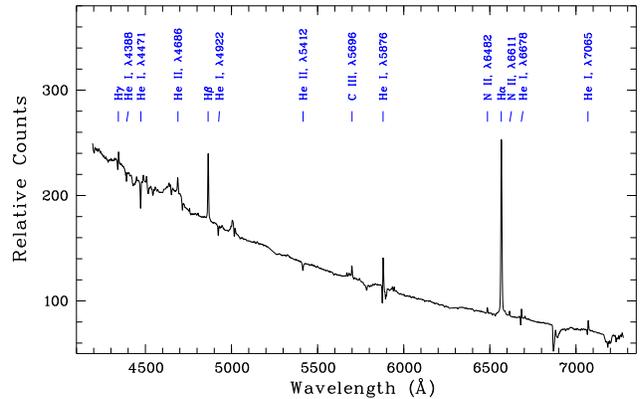}
\end{center}
\caption{The fully reduced 1D RSS spectrum of Sk$-$69$\degr$\,279 with the most prominent lines indicated.}
\label{fig:spec}
\end{figure}

\section{Observations of Sk$-$69$\degr$\,279 and its circumstellar shell}
\label{sec:obs}

\subsection{Spectroscopy}
\label{sec:spe}

Sk$-$69$\degr$\,279 and its circumstellar shell were observed on 2012 February 20 with the Southern African Large 
Telescope (SALT; Buckley, Swart \& Meiring 2006; O'Donoghue et al. 2006) using the Robert Stobie Spectrograph (RSS; 
Burgh et al. 2003; Kobulnicky et al. 2003) in the long-slit mode with a 1.25\,arcsec slit width. The slit was 
oriented in the north-south direction, i.e. at a position angle (PA) of PA=0$\degr$. The PG900 grating was used to 
cover the spectral range of 4200$-$7400~\AA \, with a final reciprocal dispersion of $0.97$~\AA \, 
pixel$^{-1}$. The spectral resolution full width at half-maximum (FWHM) is $4.45\pm$0.15~\AA. 
The total exposure time was 300~s. An Xe lamp arc spectrum was taken immediately after the 
science frames. Spectrophotometric standard stars were observed during twilight time for the
relative flux calibration. Absolute flux calibration is not feasible with SALT because the 
unfilled entrance pupil of the telescope moves during the observations. Still, we were able 
to calibrate the spectrum by adjusting its $V$ band flux to the mean $V$ magnitude of the 
star of 12.81 (see Section\,\ref{sec:phot}) using programs described in Kniazev at al. (2005).
Primary reduction of the data was done in the standard way with the SALT science pipeline 
(Crawford et al. 2010). Following long-slit data reduction was carried out in the way described 
in Kniazev et al. (2008). The resulting reduced RSS spectrum is shown in Fig.\,\ref{fig:spec}.

For spectral modelling, to search for possible spectral variability and to derive the rotational velocity, we 
obtained high-resolution spectra of Sk$-$69$\degr$\,279 with the SALT High Resolution Spectrograph (HRS; Barnes et al. 
2008; Bramall et al. 2010; Bramall et al. 2012; Crause et al. 2014) -- a dual-beam, fibre-fed \'echelle spectrograph. The data 
were taken on 2016 October 27 and 2017 September 26 in the low-resolution (LR) mode of HRS. The spectra cover a spectral range 
of 3800--9000~\AA\ at a resolution of $R=16\,000$. The exposure time in each observation was 3000\,s. 
Primary reduction of the HRS data was done using the 
SALT science pipeline (Crawford et al. 2010), which includes over-scan and gain corrections, and bias subtraction. The rest of 
the reduction was done using the {\sc midas} pipeline as described in detail in Kniazev, Gvaramadze \& Berdnikov (2016). 
Equivalent widths (EWs), FWHMs and heliocentric radial velocities (RVs) of some lines in the 2016's HRS spectrum 
(measured applying the {\sc midas} programs; see Kniazev et al. 2004 for details) are given in Table\,\ref{tab:inten}.

\begin{table}
\centering{\caption{EWs, FWHMs and RVs of some lines in the 2016's HRS spectrum of Sk$-$69$\degr$\,279. 
RVs of lines noticeably affected by P\,Cygni absorptions are starred.} 
\label{tab:inten}
\begin{tabular}{lccc} 
\hline 
 & EW($\lambda$) & FWHM($\lambda$) & RV \\
$\lambda_{0}$(\AA) Ion  & (\AA)  & (\AA) & ($\kms$) \\ 
\hline
4026\ He\ {\sc i}\          & -0.590$\pm$0.010 &  2.489$\pm$ 0.044 & 185.08$\pm$0.91   \\
4089\ Si\ {\sc iv}\         & -0.540$\pm$0.006 &  1.512$\pm$ 0.017 & 224.05$\pm$0.68   \\
4144\ He\ {\sc i}\          & -0.150$\pm$0.006 &  1.572$\pm$ 0.058 & 224.95$\pm$0.66   \\
4200\ He\ {\sc ii}\         & -0.140$\pm$0.005 &  2.336$\pm$ 0.056 & 236.94$\pm$0.61   \\
4340\ H$\gamma$\            &  0.316$\pm$0.017 &  1.097$\pm$ 0.067 & 238.44$\pm$2.04$^*$   \\
4379\ N\ {\sc iii}\         & -0.040$\pm$0.004 &  1.149$\pm$ 0.089 & 243.84$\pm$0.55   \\
4388\ He\ {\sc i}\          & -0.240$\pm$0.005 &  1.946$\pm$ 0.032 & 212.36$\pm$0.59   \\
4471\ He\ {\sc i}\          & -0.600$\pm$0.010 &  2.928$\pm$ 0.053 & 140.41$\pm$0.84$^*$   \\
4485\ S\ {\sc iv}\          &  0.140$\pm$0.004 &  1.522$\pm$ 0.025 & 233.95$\pm$0.85   \\
4504\ S\ {\sc iv}\          &  0.190$\pm$0.004 &  1.568$\pm$ 0.026 & 235.45$\pm$0.87   \\
4541\ He\ {\sc ii}\         & -0.190$\pm$0.005 &  2.163$\pm$ 0.035 & 238.44$\pm$0.56   \\
4686\ He\ {\sc ii}\         &  0.230$\pm$0.006 &  4.047$\pm$ 0.117 & 197.67$\pm$3.11   \\
4713\ He\ {\sc i}\          & -0.150$\pm$0.008 &  1.758$\pm$ 0.098 & 224.05$\pm$0.68   \\
4861\ H$\beta$\             &  2.185$\pm$0.023 &  2.301$\pm$ 0.027 & 236.04$\pm$0.83$^*$   \\
4922\ He\ {\sc i}\          & -0.260$\pm$0.007 &  2.843$\pm$ 0.073 & 185.98$\pm$0.62$^*$   \\
5016\ He\ {\sc i}\          & -0.100$\pm$0.005 &  2.399$\pm$ 0.099 &  81.95$\pm$0.51$^*$   \\
5412\ He\ {\sc ii}\         & -0.230$\pm$0.006 &  2.554$\pm$ 0.047 & 237.84$\pm$0.50   \\
5696\ C\  {\sc iii}\        &  0.380$\pm$0.005 &  1.736$\pm$ 0.013 & 227.35$\pm$0.48   \\
5876\ He\ {\sc i}\          &  1.373$\pm$0.066 &  2.201$\pm$ 0.122 & 270.52$\pm$2.68$^*$   \\
6482\ N\ {\sc ii}\          &  0.249$\pm$0.006 &  1.704$\pm$ 0.039 & 218.96$\pm$0.82   \\
6563\ H$\alpha$\            &  2.449$\pm$0.055 &  4.546$\pm$ 0.023 & 239.94$\pm$0.55   \\
6611\ N\ {\sc ii}\          &  0.319$\pm$0.005 &  1.779$\pm$ 0.023 & 218.96$\pm$0.54   \\
6678\ He\ {\sc i}\          &  0.590$\pm$0.020 &  1.816$\pm$ 0.068 & 245.94$\pm$1.34$^*$   \\
6703\ Si\ {\sc iv}\         &  0.101$\pm$0.004 &  1.625$\pm$ 0.031 & 163.19$\pm$0.66   \\
7065\ He\ {\sc i}\          &  0.638$\pm$0.020 &  1.957$\pm$ 0.071 & 236.64$\pm$1.31$^*$   \\
8502\ P16                   & -0.430$\pm$0.009 &  6.024$\pm$ 0.113 & 209.66$\pm$0.41   \\
8545\ P15                   & -0.410$\pm$0.018 &  5.725$\pm$ 0.272 & 221.65$\pm$0.67   \\
8598\ P14\                  & -0.510$\pm$0.008 &  5.716$\pm$ 0.064 & 222.25$\pm$0.36   \\
8665\ P13\                  & -0.700$\pm$0.010 &  5.880$\pm$ 0.067 & 228.25$\pm$0.41   \\
8750\ P12\                  & -0.750$\pm$0.012 &  6.299$\pm$ 0.101 & 221.05$\pm$0.48   \\
\hline
\end{tabular}
  }
\end{table}

To investigate Sk$-$69$\degr$\,279 in the UV range, we retrieved its spectra, obtained with the 
Cosmic Origins Spectrograph (COS) on board the {\it Hubble Space Telescope} ({\it HST}), from 
the Mikulski Archive for Space Telescopes (MAST)\footnote{https://archive.stsci.edu/}. The spectra 
were obtained with the G130M and G160M gratings covering the spectral range from 1170 to 1770\,\AA \, 
with a resolution of $R=16\,000-21\,000$. 

\begin{table}
  \caption{Photometry of Sk$-$69$\degr$\,279.}
  \label{tab:phot}
  \renewcommand{\footnoterule}{}
  \begin{center}
  \begin{minipage}{\textwidth}
    \begin{tabular}{lccc}
      \hline
      Date & $B$ & $V$ & $I_{\rm c}$ \\
      \hline
      2013 January 3 & 12.86$\pm$0.01 & 12.81$\pm$0.01 & 12.67$\pm$0.01 \\                     
      2014 April 5 & 12.92$\pm$0.02 & 12.83$\pm$0.01 & 12.75$\pm$0.01 \\
      2016 March 24 & 12.82$\pm$0.01 & 12.79$\pm$0.01 & 12.66$\pm$0.01 \\
      \hline
    \end{tabular}
    \end{minipage}
    \end{center}
   \end{table}

\begin{figure}
\begin{center}
\includegraphics[width=6cm,angle=270,clip=]{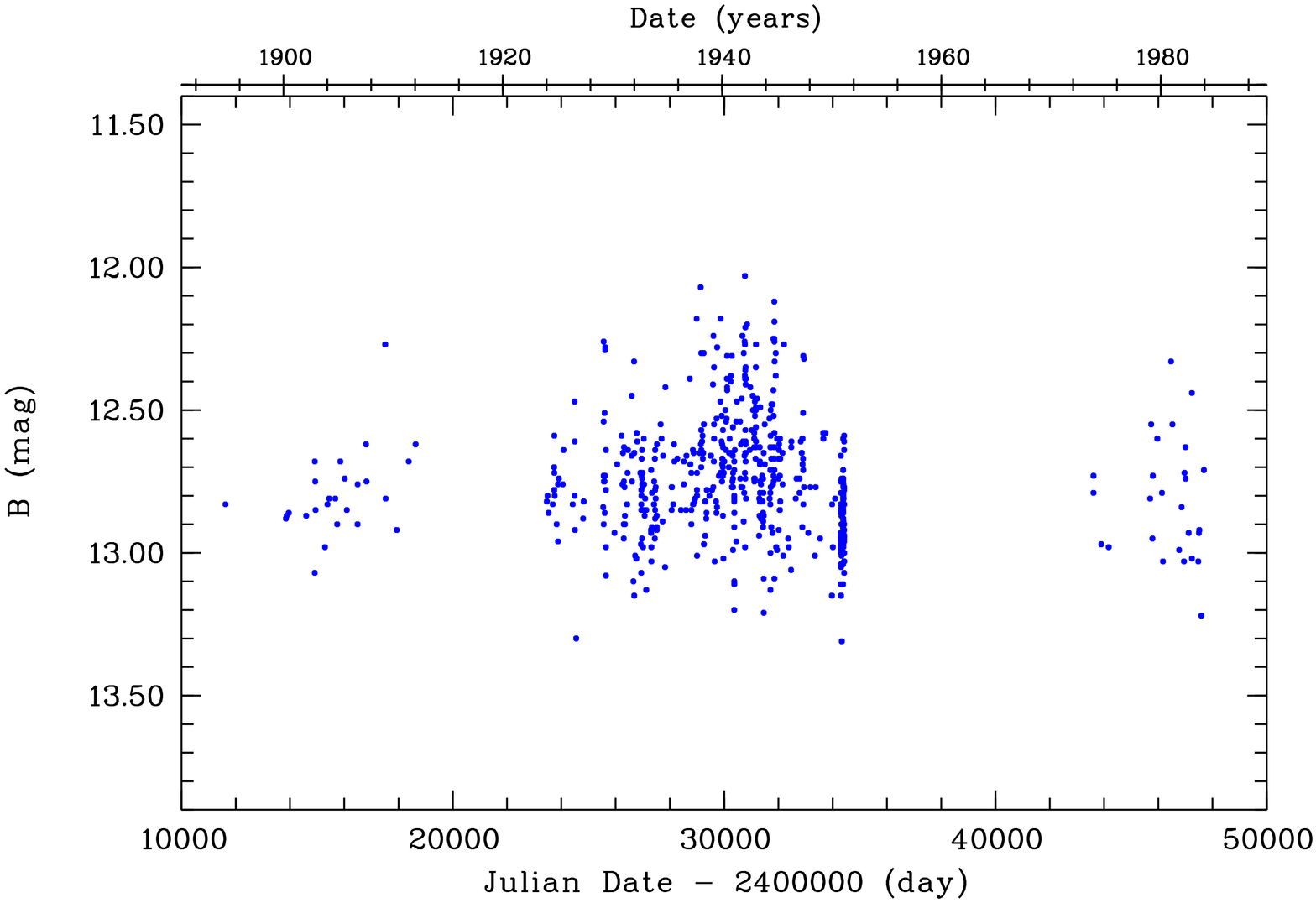}
\includegraphics[width=6cm,angle=270,clip=]{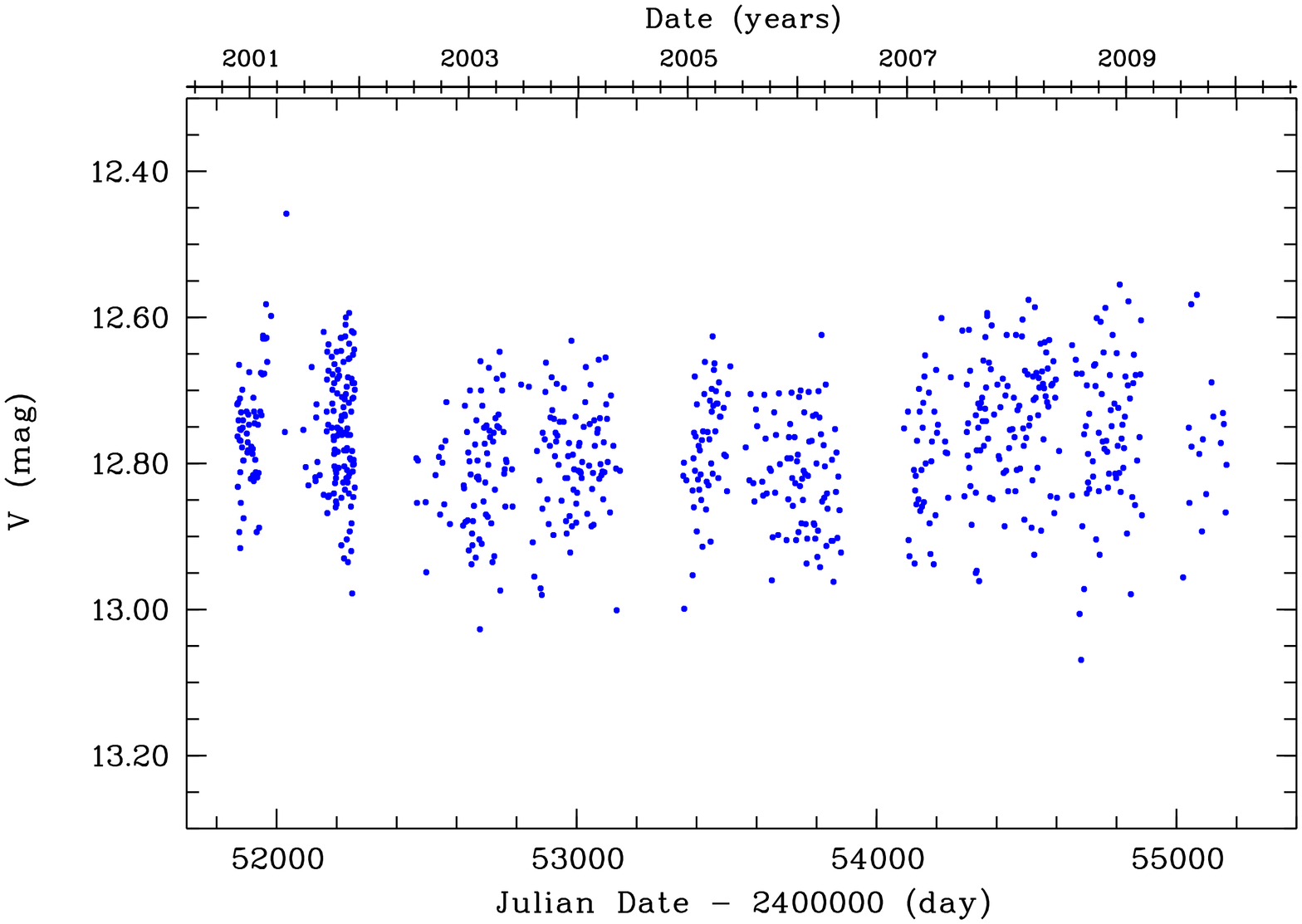}
\includegraphics[width=6cm,angle=270,clip=]{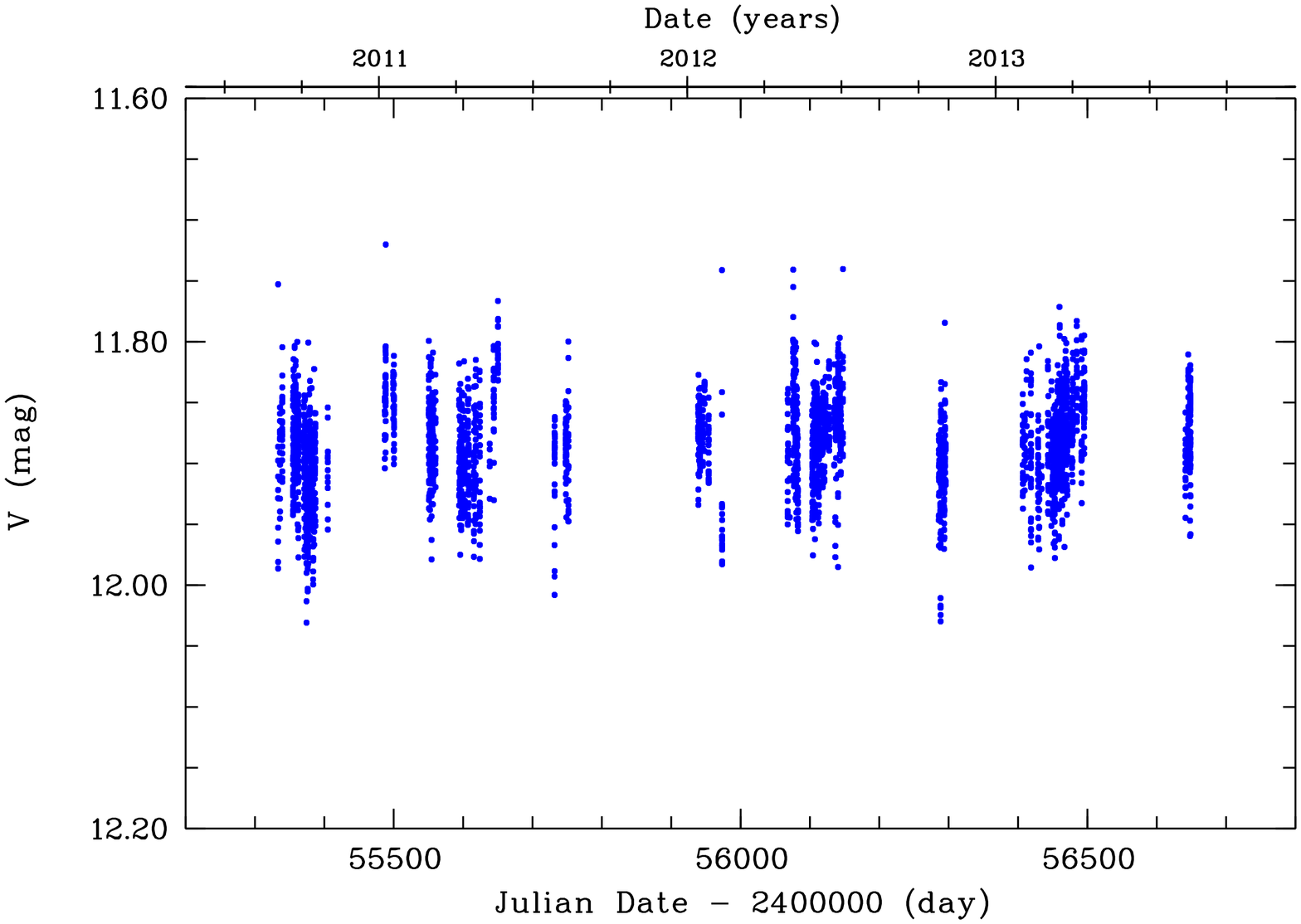}
\end{center}
\caption{The Harvard DASCH (upper panel), ASAS (middle panel) and OMC-INTEGRAL (bottom panel) light 
curves of Sk$-$69$\degr$\,279. Note that the bottom light curve systematically overestimates the 
brightness of Sk$-$69$\degr$\,279 because of the large OMC photometric aperture (see the text for 
details).}
\label{fig:phot}
\end{figure}

\subsection{Photometry}
\label{sec:phot}

To search for possible photometric variability of Sk$-$69$\degr$\,279, we determined its $B, V$ 
and $I_{\rm c}$ magnitudes on CCD frames obtained with the 76-cm telescope of the South African 
Astronomical Observatory during our threee observing runs in 2013--2016. We used an SBIG ST-10XME 
CCD camera equipped with $BVI_{\rm c}$ filters of the Kron-Cousins system (see e.g. Berdnikov et al. 
2012). The resulting photometry is presented in Table\,\ref{tab:phot}. From table it follows that  
Sk$-$69$\degr$\,279 did not experience major changes in its brightness during the three years: its
$B$, $V$ and $I_{\rm c}$ magnitudes remained almost constant with mean values of 12.87$\pm$0.01, 
12.81$\pm$0.01 and 12.69$\pm$0.01 mag, respectively. The $V$ magnitude and the mean value of the 
$B-V$ colour of 0.06$\pm$0.01 are in good agreement with the values measured for Sk$-$69$\degr$\,279 
by Isserstedt (1975), respectively, 12.79 and 0.05 mag.

Also, we used data from the archives of the Digital Access to a Sky Century @ Harvard (DASCH; Grindlay 
et al. 2009) project, the All-Sky Automated Survey (ASAS; Pojma\'nski 2002) and the Optical Monitoring 
Camera (OMC) on board the International Gamma-Ray Astrophysics Laboratory (INTEGRAL) satellite 
(Alfonso-Garz\'on et al. 2012). The light curves based on these data are shown in Fig.\,\ref{fig:phot}.
The DASCH provides 1484 measurements of the $B$ magnitude of Sk$-$69$\degr$\,279 over the time interval 
of $\approx$1890--1985 with a gap between 1955 and 1970. After rejection of low quality data points, we
are left with 553 measurements with a median value of 12.76$\pm$0.22 mag,  
which agrees with the mean $B$ magnitude of 12.87$\pm$0.01, based on our CCD observations. Similarly, the 
ASAS light curve in the $V$-band (based on 733 measurements) shows that the brightness of the star in 
2001--2009 varies by $\approx\pm$0.1 mag around the median value of $\approx$12.8 mag. The OMC light curve 
of Sk$-$69$\degr$\,279 (based on 2979 measurements) also shows that the $V$-band brightness varies by 
$\approx\pm$0.1 mag. The OMC photometry, however, is contaminated by nearby stars within the 52.5$\times$52.5 
arcsec photometric aperture, which results in the systematic increase of the $V$ magnitude by $\approx$0.9 mag.

\begin{figure*}
{\centering \resizebox*{1.7\columnwidth}{!}{\includegraphics[angle=0]{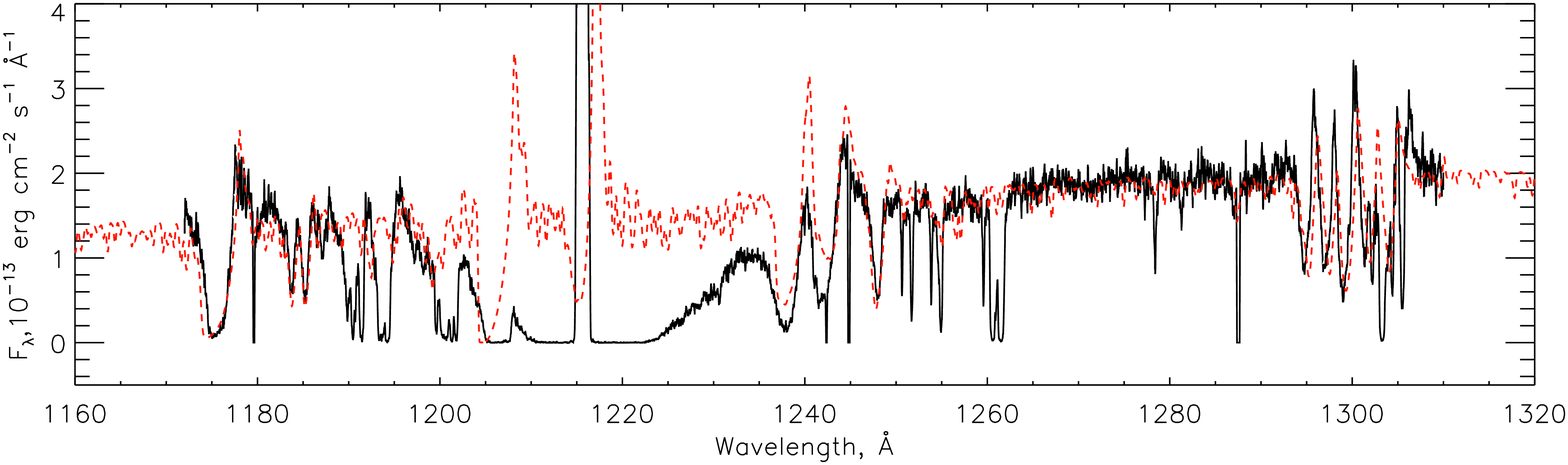}}}\\
{\centering \resizebox*{1.7\columnwidth}{!}{\includegraphics[angle=0]{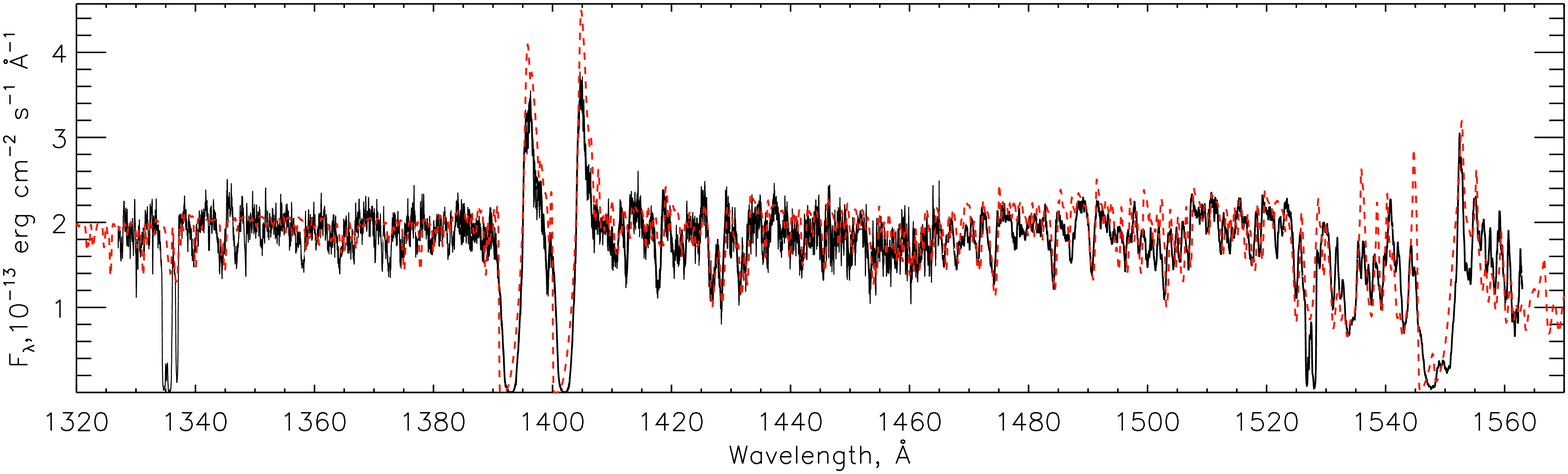}}}\\
{\centering \resizebox*{1.7\columnwidth}{!}{\includegraphics[angle=0]{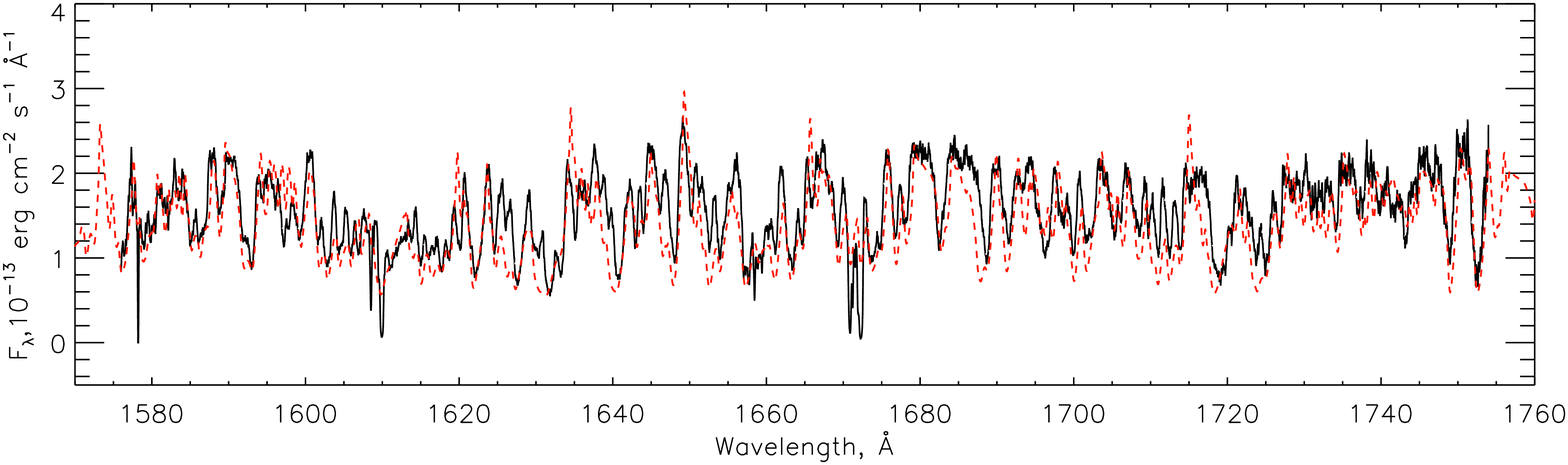}}}\\\
 \caption{Comparison of the HST/COS UV spectrum (black solid line) with the model one (red dashed line),
 scaled to the distance to LMC and reddened by $E(B-V)=0.31$ mag. 
}
\label{fig:UV}
\end{figure*}

\begin{figure*}
{\centering \resizebox*{1.7\columnwidth}{!}{\includegraphics[angle=0]{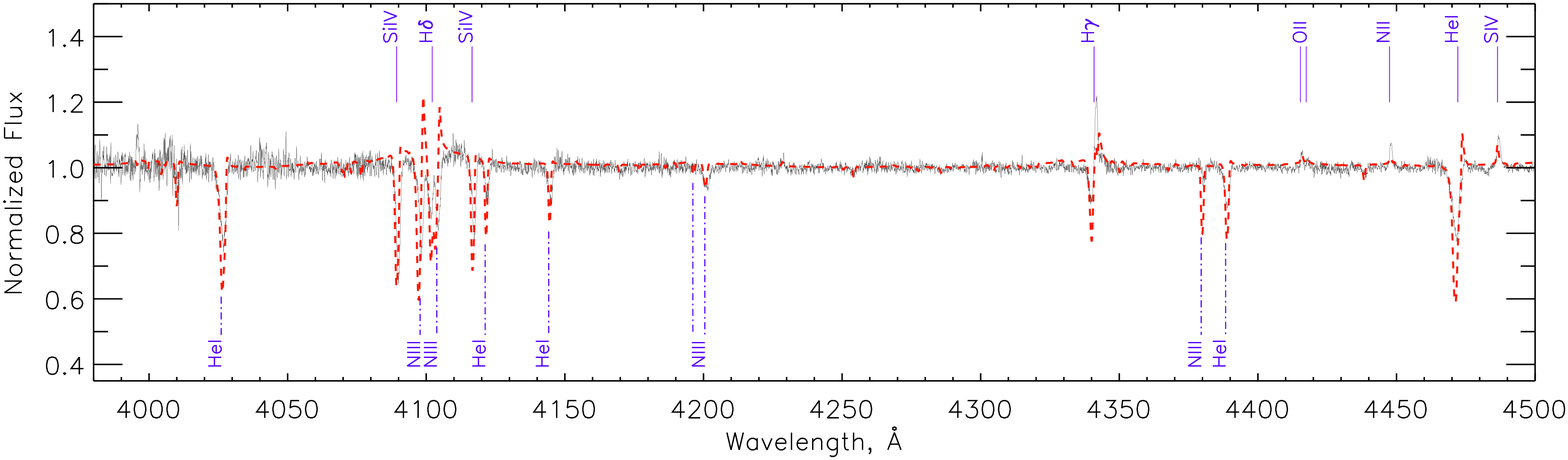}}}\\
{\centering \resizebox*{1.7\columnwidth}{!}{\includegraphics[angle=0]{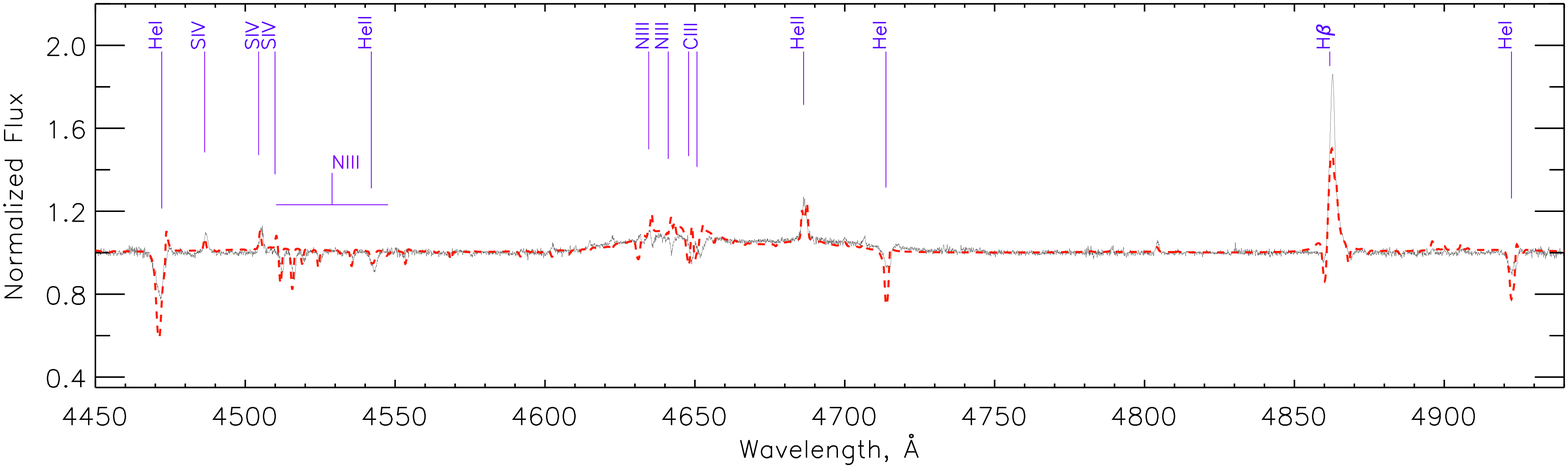}}}\\
{\centering \resizebox*{1.7\columnwidth}{!}{\includegraphics[angle=0]{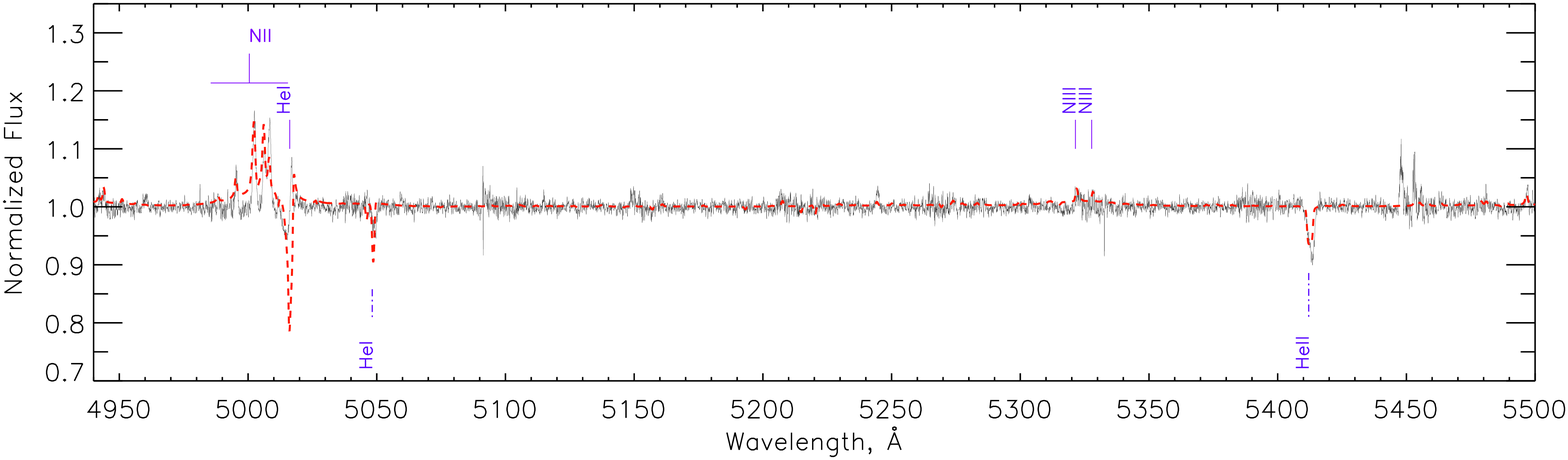}}}\\
\caption{Normalized HRS spectrum of Sk$-$69$\degr$\,279 taken on 2016 October 27 (black solid line), compared with 
the best-fitting {\sc cmfgen} model (red dashed line) with the parameters as given in Table\,\ref{tab:par}. 
}
\label{fig:model}
\end{figure*}

\addtocounter{figure}{-1}
\begin{figure*}
{\centering \resizebox*{1.7\columnwidth}{!}{\includegraphics[angle=0]{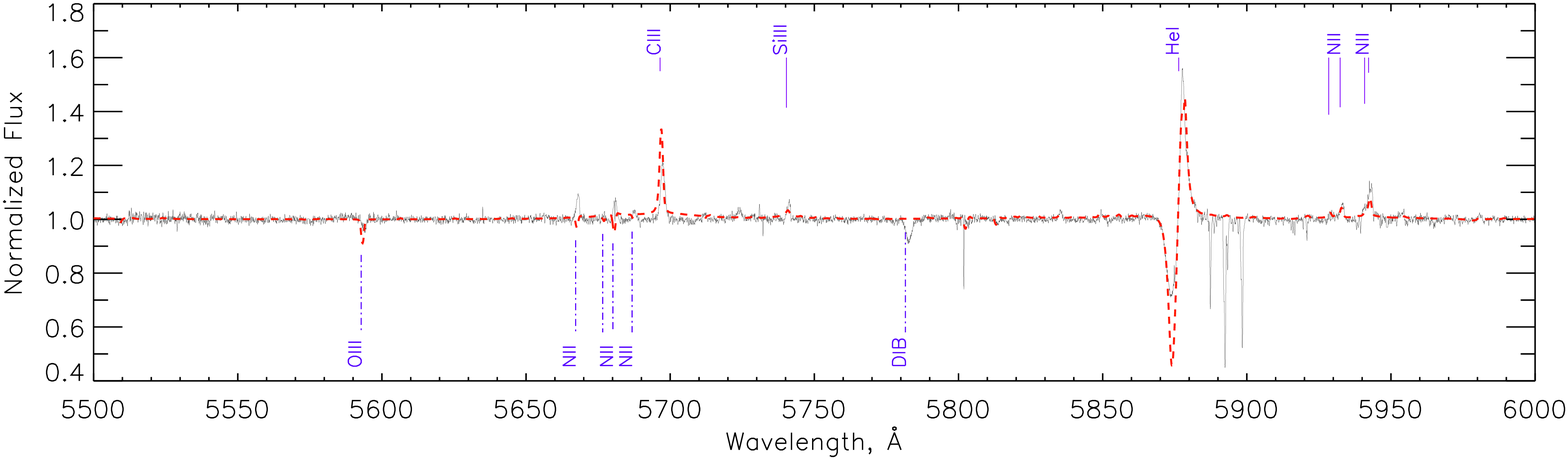}}}\\
{\centering \resizebox*{1.7\columnwidth}{!}{\includegraphics[angle=0]{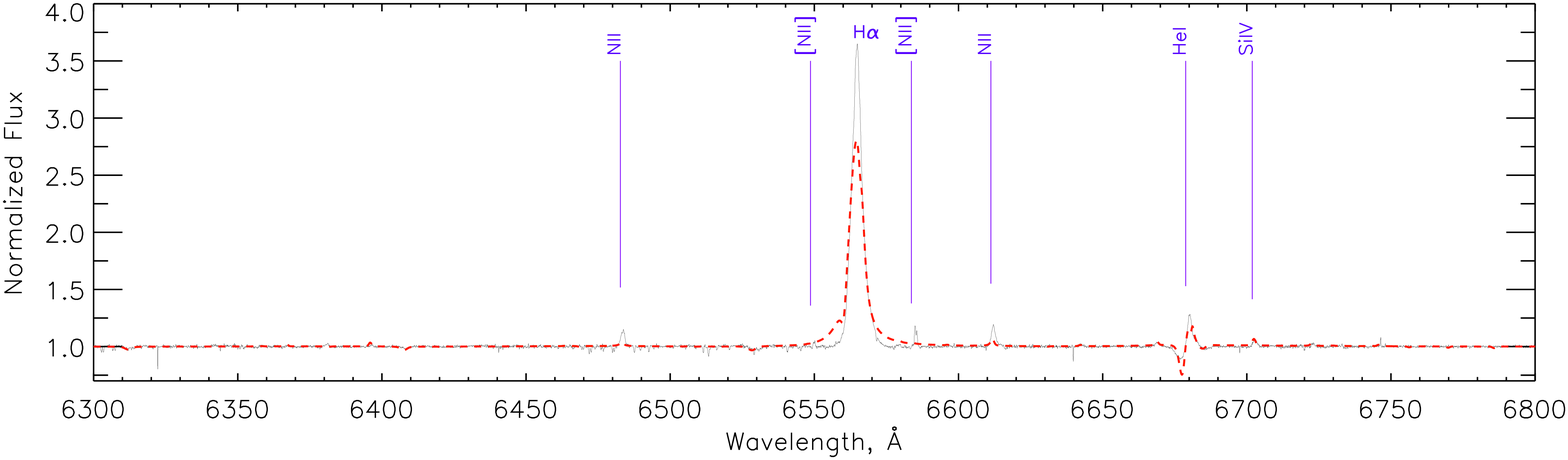}}}\\
{\centering \resizebox*{1.7\columnwidth}{!}{\includegraphics[angle=0]{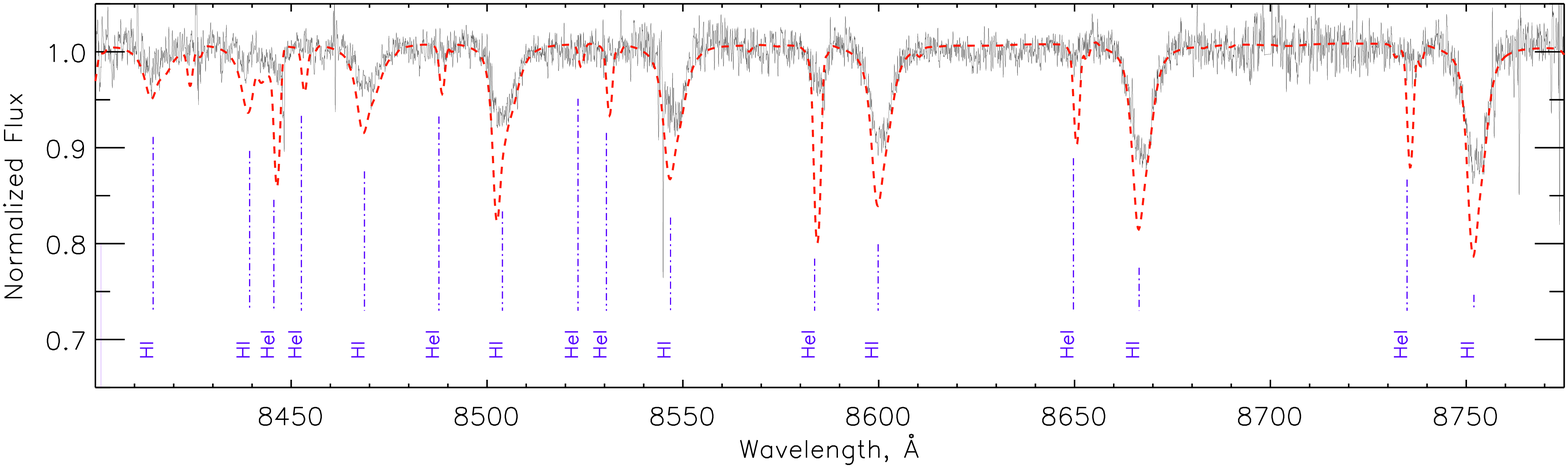}}}\\
\caption{(Continued)}
\label{fig:model}
\end{figure*}

\section{Sk$-$69$\degr$\,279: spectral analysis and stellar parameters}
\label{sec:spec}

\subsection{Classification of Sk$-$69$\degr$\,279}
\label{sec:class}

Fig.\,\ref{fig:spec} shows that the spectrum of Sk$-$69$\degr$\,279 is dominated by emission lines of H, 
He\,{\sc i}, He\,{\sc ii}, N\,{\sc ii} and C\,{\sc iii}, some of which show P\,Cygni profiles. Further 
emission lines in the spectrum (see Fig.\,\ref{fig:model}) are permitted lines of O\,{\sc  ii}, N\,{\sc 
iii}, C\,{\sc iii}, Si\,{\sc iii}-{\sc iv}, and C\,{\sc iv}. 

To classify Sk$-$69$\degr$\,279, we use the classification criteria of Conti \& Alschuler (1971). Namely, 
the logarithm of the ratio of the EWs of the He\,{\sc i} $\lambda$4471 and He\,{\sc ii} $\lambda$4541 
lines (see Table\,\ref{tab:inten}), $\log$(EW$_{4471}$/EW$_{4541}$)=0.50$\pm$0.02, indicates that 
Sk$-$69$\degr$\,279 is an O9.5 star (see table\,3 in Conti \& Alschuler 1971). The more recent 
classification criteria of Sota et al. (2014; see their table\,3) based on the ratios of peak intensities 
of absorption-line pairs, $I$(He\,{\sc ii} $\lambda$4541)/$I$(He\,{\sc i} $\lambda$4388)=0.8 and 
$I$(He\,{\sc ii} $\lambda$4200)/$I$(He\,{\sc i} $\lambda$4144)=0.9 (both slightly less than 1), and 
$I$(Si\,{\sc iii} $\lambda$4552)/$I$(He\,{\sc ii} $\lambda$4541)$\ll$1, suggest that Sk$-$69$\degr$\,279 
is an O9.2 star. The same spectral type also follows from the classification scheme for O stars based on the 
ratio of EWs of the lines He\,{\sc i} $\lambda$4922 and He\,{\sc ii} $\lambda$5411 (Kerton, Ballantyne \& 
Martin 1999):
\begin{eqnarray}
{\rm SpT}=(9.04\pm0.10)+(4.10\pm0.23)\log({\rm EW}_{4922}/{\rm EW}_{5411}) \, . \nonumber 
\end{eqnarray}
Using this equation and EWs from Table\,\ref{tab:inten}, one finds SpT=9.26.

To determine the luminosity class of Sk$-$69$\degr$\,279, we use table\,5 in Conti \& Alschuler (1971) and 
EWs of the Si\,{\sc iv} $\lambda$4089 and He\,{\sc i} $\lambda$4144 lines from Table\,\ref{tab:inten}. With 
$\log$(EW$_{4089}$/EW$_{4144}$)=0.56$\pm$0.02, one finds that Sk$-$69$\degr$\,279 is a supergiant, which 
agrees with location of this star in the LMC (cf. Nandy et al. 1984). Similarly, using the dependence of 
EW of the S\,{\sc iv}\,$\lambda$4483 emission line on the luminosity class of O stars (see fig.\,4 in Morrell, 
Walborn \& Fitzpatrick 1991) and the measured EW$_{4483}$=140 m\AA \, (see Table\,\ref{tab:inten}), one finds 
that the luminosity class of Sk$-$69$\degr$\,279 is either Ia or Iab, while the ratio $I$(Si\,{\sc iv} 
$\lambda$4089)/$I$(He\,{\sc i} $\lambda$4026) of $<1$ allows one to narrow down the luminosity class to Ia 
(see table\,6 in Sota et al. 2011).

The presence of the He\,{\sc ii} $\lambda$4686 and N\,{\sc iii} $\lambda\lambda$4634--40--42 emission lines 
indicates that Sk$-$69$\degr$\,279 is an Of star (e.g. Sota et al. 2011), we therefore add a suffix `f' to the 
spectral subtype, which then becomes O9.2\,Iaf.

Using $B-V=0.06\pm0.01$ (see Section\,\ref{sec:phot}) and the intrinsic colour of late O supergiants of 
$(B-V)_0=-0.26$ mag (Martins \& Plez 2006), one finds the colour excess of $E(B-V)=0.32\pm0.01$ mag towards 
Sk$-$69$\degr$\,279, which is in a good agreement with independent estimates derived in Sections\,\ref{sec:mod} 
and \ref{sec:shell}. With the distance modulus for the LMC of 18.49 and assuming $A_V=3.1E(B-V)$ (cf. Howarth 
1983), one obtains the absolute visual magnitude $M_V=-6.67\pm0.03$ mag. Then, with $T_{\rm eff}=24$\,kK 
(see Table\,\ref{tab:par} in the next section) and BC=$-2.34\pm0.05$ mag (Crowther, Lennon \& Walborn 2006), 
one finds $M_{\rm bol}=-9.01\pm0.06$ mag and $\log(L_*/\lsun)=5.50\pm0.02$, which well agrees with the results 
of spectral modelling.

\subsection{Spectral modelling}
\label{sec:mod}

To determine the stellar parameters we have used the {\sc cmfgen} model atmosphere 
code (Hillier \& Miller 1998). This code solves the radiative transfer 
equations for objects with spherically symmetric extended outflows using either 
the Sobolev approximation or the full comoving-frame solution of the radiative 
transfer equation. {\sc cmfgen} incorporates line blanketing, the effect of Auger 
ionization and clumping. Every model is defined by the hydrostatic stellar radius 
$R_*$, bolometric luminosity $L_*$, mass-loss rate $\dot{M}$, volume filling factor $f$, 
wind terminal velocity $v_\infty$, stellar mass $M_*$, and abundances of included chemical 
elements $Z_i$. Our calculations included H, He, C, N, O, Si, S, P and Fe. The mass-loss 
rate, density, and velocity are related to each other via the continuity equation. The 
best-fitting model is shown in Fig.\,\ref{fig:UV} (UV range) and Fig.\,\ref{fig:model} (optical 
range)\footnote{Note that the [N\,{\sc ii}]~$\lambda\lambda 6548,84$ lines visible in 
the HRS spectrum are of nebular origin (see Section\,\ref{sec:shell}).}.

As an initial photospheric density structure we adopted the {\sc tlusty} hydrostatic model atmosphere 
of Hubeny \& Lanz (1995) and Lanz \& Hubeny (2003) for deep quasi-static layers, which are connected to 
the wind with standard $\beta$-velocity law just above the sonic point. The $\beta$-velocity law 
is one of the basic simplifications typically adopted when constructing atmospheric models of hot stars. 
The radiation-driven wind theory (Puls et al. 1996; Lamers \& Cassinelli 1999) predicts values of 
$\beta=0.5-1$. Our study, however, suggests that a model with $\beta$=3 and the connection point located 
at approximately $40 \, \kms$ describes the observed spectrum better than models with other values of $\beta$
and the connection point at 10 or $5 \, \kms$. Although the values as high as $\beta=3-3.5$ disagree with 
the theoretical predictions, they were often employed in modelling of (c)LBVs and blue supergiants (e.g., 
Kudritzki et al. 1999; Groh et al. 2009; Evans et al. 2004; Mahy et al. 2016). One of the first discussions 
of the discrepancy between the modelling results and the theoretical predictions is given in Hillier et al. 
(2003).

For measuring the stellar temperature $T_*$ (determined at the radius where the Rosseland optical 
depth is 20) and the effective temperature $T_{\rm{eff}}$ (determined at the radius at which the 
Rosseland optical depth is 2/3) with {\sc cmfgen}, we compared the intensities of different 
ion lines (He\,{\sc i}, {\sc ii}; C\,{\sc iii}, {\sc iv}; N\,{\sc ii}, {\sc iii}; 
Si\,{\sc iii}, {\sc iv}) in the HRS spectrum, i.e., we used the traditional ionization-balance method.

To determine the luminosity of Sk$-$69$\degr$\,279, we recomputed the fluxes 
for the distance to the LMC. The resulting fluxes were corrected for the interstellar 
extinction, and then compared to the calculated spectra convolved with the transmission curves 
of the standard $B$, $V$ and $I_{\rm c}$ filters.

\begin{figure*}
\begin{center}
\includegraphics[width=14cm,angle=0,clip=]{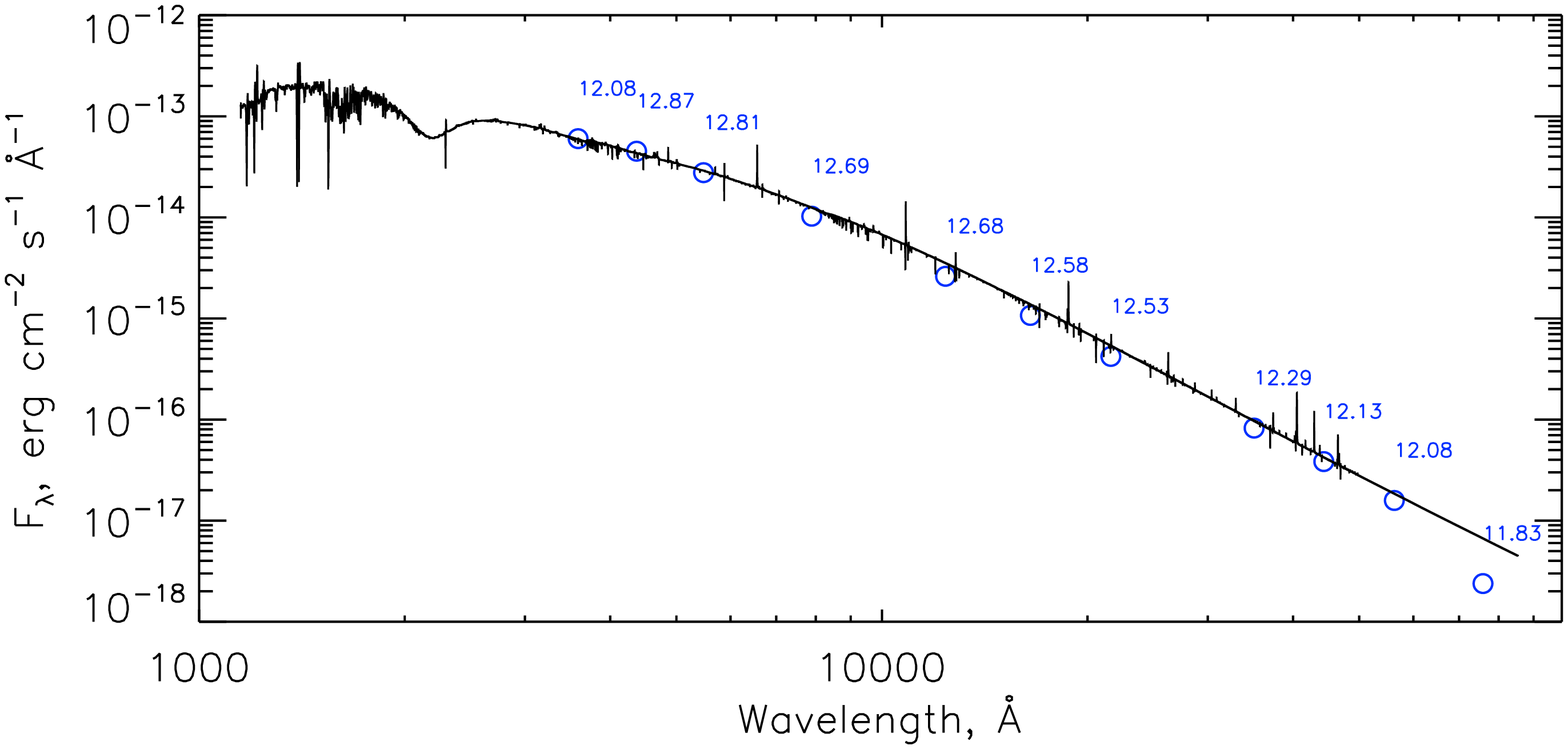}
\end{center}
\caption{Observed flux distribution of Sk$-$69$\degr$\,279 in absolute units, based on the photometric 
measurements (blue circles) compiled in Table\,\ref{tab:phot}, compared to the emergent flux of the 
reddened model spectrum (black/noisy line) with the parameters as given in Table\,\ref{tab:par}.}
\label{fig:sed}
\end{figure*}

The colour excess $E(B-V)$ towards Sk$-$69$\degr$\,279 was estimated by comparing the 
spectral energy distribution (SED) in the model spectrum with the observed UV and HRS
spectra and photometric measurements compiled in Table\,\ref{tab:phot}.
We found that the slopes of the model and observed spectra match each other if $E(B-V)=0.31$\,mag 
(Fig.~\ref{fig:sed}). This value is within the error margins of $E(B-V)=0.34\pm0.02$ mag 
derived from the Balmer decrement in the spectrum of the circumstellar shell (see 
Section\,\ref{sec:shell}).

\begin{figure*}
{\centering \resizebox*{1.7\columnwidth}{!}{\includegraphics[angle=0]{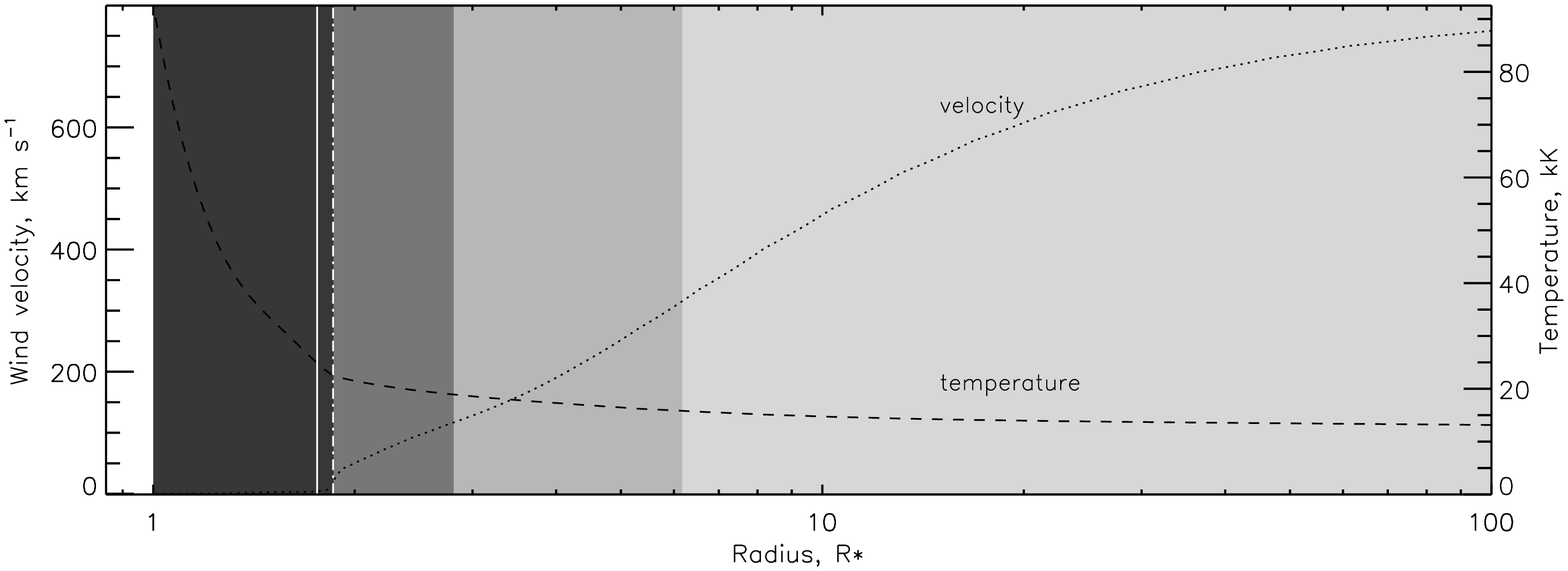}}}\\
{\centering \resizebox*{1.7\columnwidth}{!}{\includegraphics[angle=0]{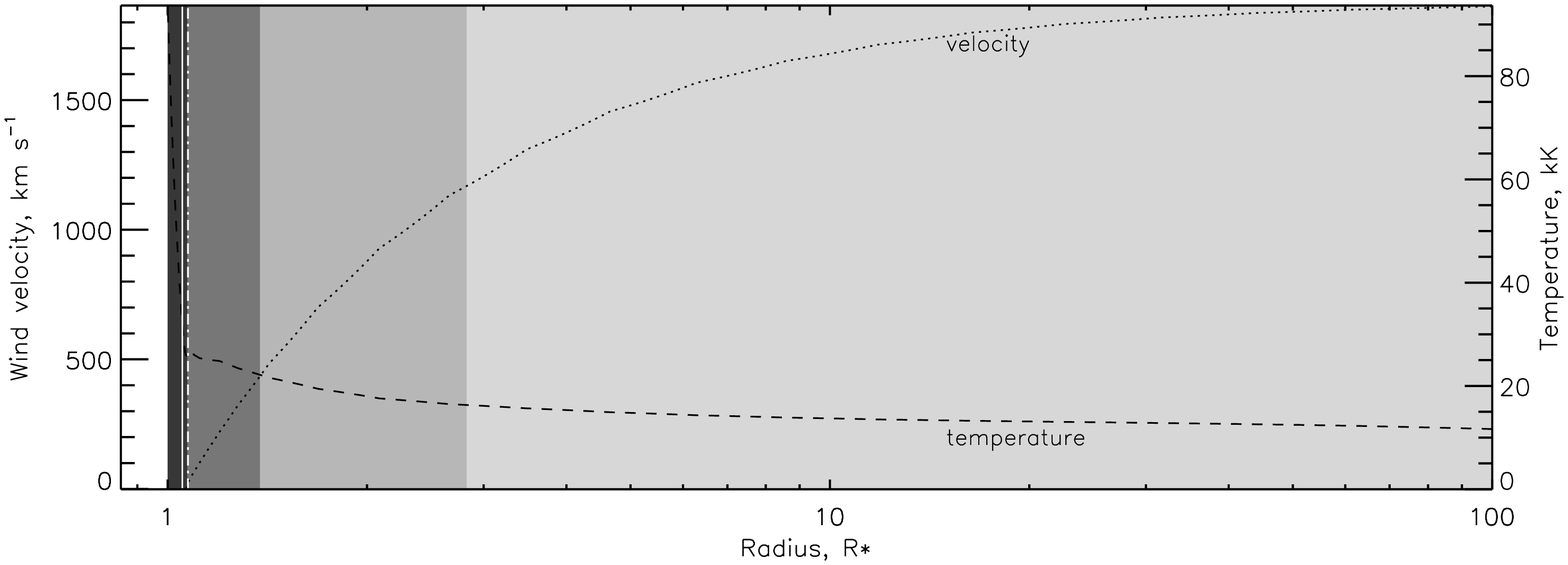}}}\\
\caption{The wind structure of the best-fit {\sc cmfgen} model atmosphere of Sk$-$69$\degr$\,279 (upper panel) 
and a model atmosphere of an O9 supergiant (Martins et al. 2005). The differently coloured regions 
correspond, from left to right, to densities n$_{\rm e}>10^{12} \, {\rm cm}^{-3}$, $10^{12}>n_{\rm e}>10^{11} \, 
{\rm cm}^{-3}$, $10^{11}>n_{\rm e}>10^{10} \, {\rm cm}^{-3}$, $10^{10}>n_{\rm e}>10^9 \, {\rm cm}^{-3}$, and 
$n_{\rm e}<10^9$. The white vertical solid and dot dashed lines correspond, respectively, to a radius at which the 
Rosseland optical depth equals 2/3 and to a sonic radius. Black dotted and dashed lines plot, respectively, the 
wind velocity and temperature profiles as a function of radius (on a logarithmic scale).}
\label{fig:wind}
\end{figure*}

As a next step, $v_\infty$ and $\dot{M}$ were estimated by using lines with P\,Cygni  
profiles and intensities of the emission lines. Our modelling also included the clumping 
described by the volume filling factor $f=\bar{\rho}/\rho(r)$, where $\bar{\rho}$ is the homogeneous 
(unclumped) wind density and ${\rho}$ is the density in clumps. It is assumed that the clumps are
optically thin, while the interclump medium is void (Hillier \& Miller 1999).

The filling factor depends on the radius as $f(r)=f_\infty +(1-f_\infty)\exp(-v(r)/v_{\rm cl})$, where 
$f_\infty$ describes the density contrast and $v_{\rm cl}$ is a characteristic velocity at which clumping 
starts to be important. \v{S}urlan et al. (2012, 2013) demonstrated that the clumping starts in the innermost 
layers of the wind, therefore we adopted $v_{\rm cl}=5 \, \kms$ and determined $f_\infty=0.5$ by fitting 
emission lines. Puls et al. (2006) and Najarro, Hanson \& Puls (2011) studied
the radial distribution of the clumping by simultaneous modelling of H$\alpha$, infrared, millimetre, 
and radio observations, and demonstrated that it decreases at large radii. And indeed, our computations 
show that, after adjusting all other parameters, the profiles of the H$\alpha$ and H$\beta$ lines are reproduced 
better if the clumping starts to disappear at velocities greater than $700 \, \kms$.

The simultaneous presence of `hot' (e.g. He\,{\sc ii} $\lambda\lambda$4686, 5411) and `cold' (e.g. N\,{\sc ii} 
$\lambda\lambda$6482, 6611) lines in the spectrum suggests that Sk$-$69$\degr$\,279 has an extended atmosphere 
with a large difference between $T_*$ and $T_{\rm eff}$. In Fig.\,\ref{fig:wind}, we plot the density, temperature 
and velocity profiles of our best-fit model atmosphere corresponding to $T_*\approx30$\,kK and $T_{\rm eff}\approx24$ 
kK (upper panel) and the model atmosphere of a `normal' O9 supergiant (lower panel; Martins, Schaerer \& Hillier 
2005). Comparison of the panels shows that Sk$-$69$\degr$\,279 has a more inflated atmosphere and a much
slower wind velocity as compared with blue supergiants of similar spectral type (e.g. Markova et al. 2003; Mokiem et al. 2007).
This could be connected with the bi-stability jump mechanism (Pauldrach \& Puls 1990; Lamers \& Pauldrach 1991), which 
is manifested in a factor of 2 decrease in $v_\infty$ when $T_{\rm eff}$ drops below a critical value of $\approx21-25$\,kK 
(Vink, de Koter \& Lamers 1999), and in a drastic increase in $\dot{M}$ (Lamers, Snow \& Lindholm 1995; Vink et al. 1999), 
accompanied by the formation of an extended atmosphere (Smith, Vink \& de Koter 2004).

With the 2016's HRS spectrum, we also estimated the projected rotational and heliocentric radial velocities of 
Sk$-$69$\degr$\,279. To determine the projected rotational velocity, $v\sin i$, we used the correlations between this 
velocity and FWHMs of the He\,{\sc i} $\lambda\lambda$4026, 4144, 4388 and 4471 lines; see equations (1)--(4) in 
Steele, Negueruela \& Clark (1999). For FWHM(4026)=2.49$\pm$0.04 \AA, FWHM(4144)=1.57$\pm$0.06 \AA, 
FWHM(4388)=1.95$\pm$0.03 \AA \, and FWHM(4471)=2.93$\pm$0.05 \AA, we found $v\sin i\approx114$, 70, 82 and $121 \, \kms$, 
respectively, with a mean value of $97\pm10 \, \kms$ (here we conservatively set the margins of error of $10 \, \kms$). 
This figure should be considered as an upper limit to the projected rotational velocity because the lines might be broadened 
because of macroturbulence. Note that a similar value of $v\sin i$ of $84\pm10 \, \kms$ was derived by Penny \& Gies (2009) 
from UV lines in the {\it Far Ultraviolet Spectrographic Explorer} ({\it FUSE}) spectrum of Sk$-$69$\degr$\,279. 

The heliocentric radial velocity of Sk$-$69$\degr$\,279 was obtained using RVs of lines not affected by P\,Cygni absorptions. 
We found the mean value of RV of $226\pm3 \, \kms$, which agrees with the systemic velocity of the shell (see 
Section\,\ref{sec:shell}). To search for possible radial velocity variability, we carried out a cross-correlation analysis 
on the two HRS spectra in the spectral range of 8460--8765~\AA \,, which is dominated by Paschen absorption lines. We found 
that in the second spectrum these lines become shifted bluewards by $16.3\pm0.2 \, \kms$. Although this change in the radial
velocity might be caused by the duplicity of the star, the more likely explanation is that it is due to the 
photospheric variability, which is typical of luminous blue supergiants. For example, similar radial velocity variations 
were detected in the spectra of two blue supergiants with bipolar circumstellar nebulae, Sher\,25 (Smartt et al. 2002; 
Taylor et al. 2014) and HD\,168625 (Mahy et al. 2016), both of which are considered to be cLBVs.

\begin{figure}
\begin{center}
\includegraphics[width=6cm,angle=270,clip=]{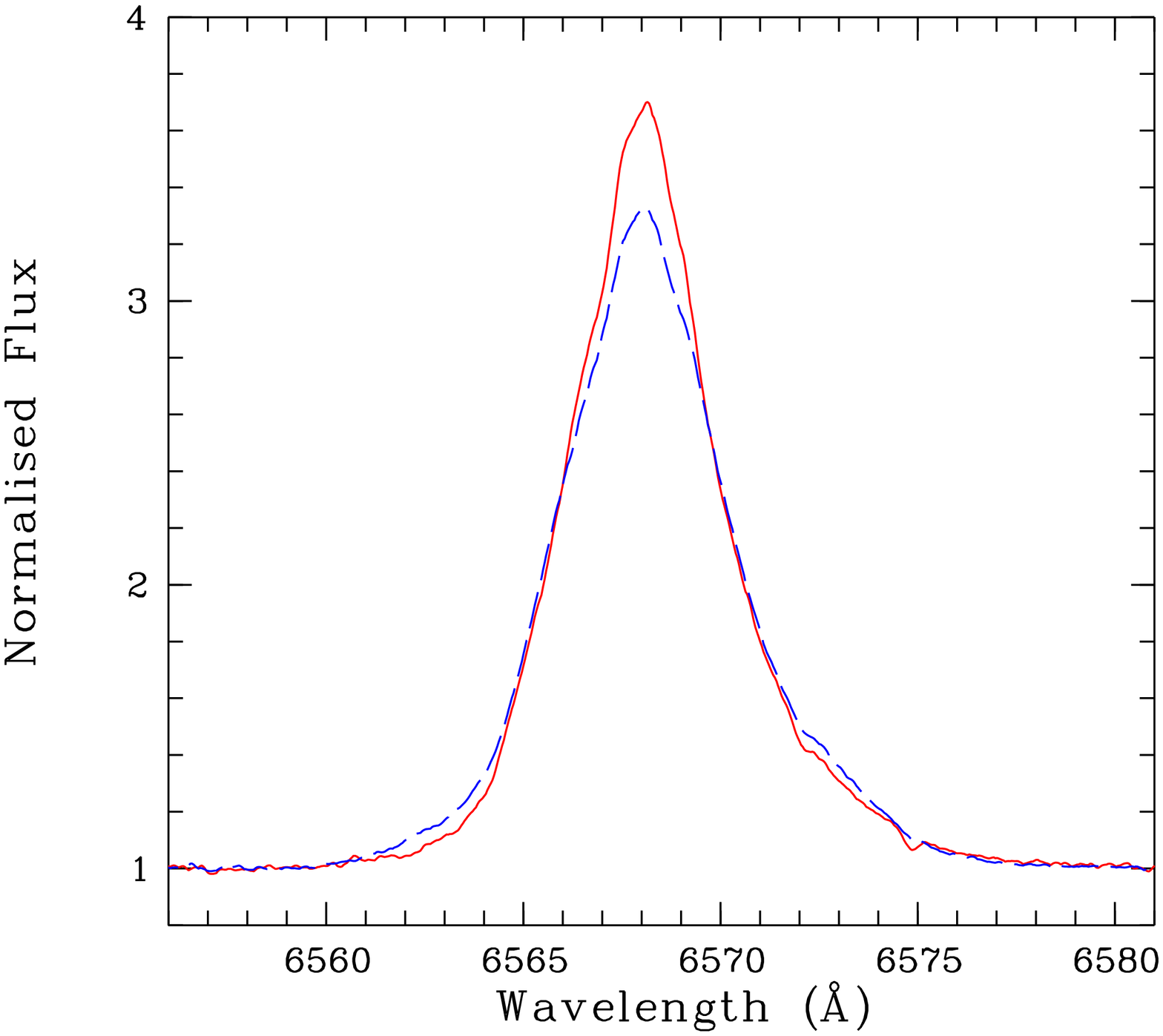}
\includegraphics[width=6cm,angle=270,clip=]{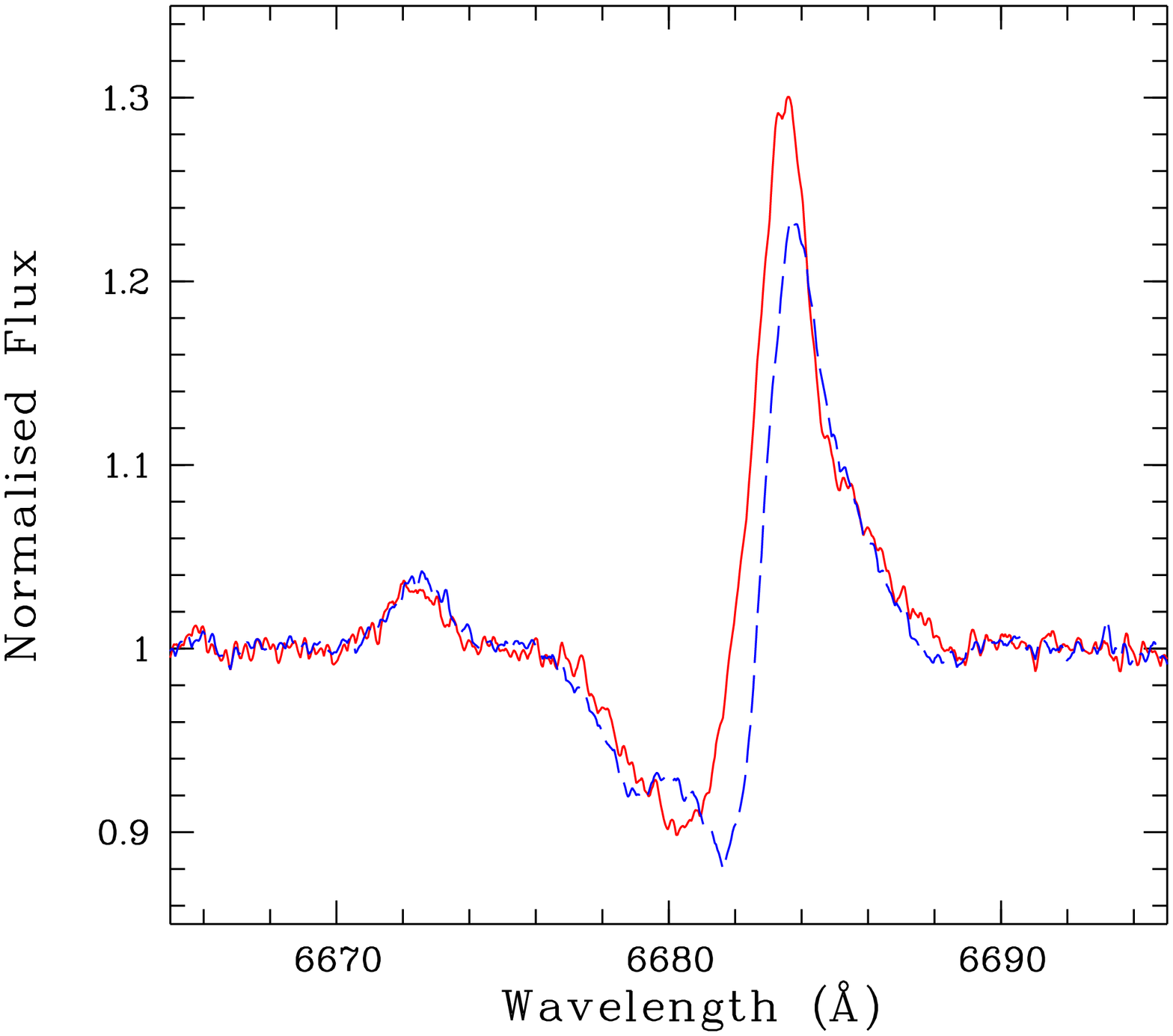}
\end{center}
\caption{Changes in the H$\alpha$ (upper panel) and He\,{\sc i} $\lambda$6678 (bottom panel) line profiles: 
2016 October 27 (red solid line) and 2017 September 26 (blue dashed line).}
\label{fig:var}
\end{figure}

Comparison of the HRS spectra revealed further signatures of the wind variability. Namely, we found that i) the intensity of the
H$\alpha$ and H$\beta$ lines has decreased by about 10 per cent in 2017, while the RVs of these lines remain the same 
(Fig.\,\ref{fig:var}), ii) the He\,{\sc ii} $\lambda\lambda$4541, 4686 and 5412 lines become shifter redwards, and iii) P\,Cygni
profiles of He\,{\sc i} lines (e.g. He\,{\sc i} $\lambda\lambda$4471, 4921, 5015, 5876 and 6678) show changes in their red wings, 
while the blue ones remain intact (see Fig.\,\ref{fig:var}).

A summary of the parameters of Sk$-$69$\degr$\,279 derived above is given in Table~\ref{tab:par}.
To this table we also added surface abundances of the basic elements in the atmosphere of 
Sk$-$69$\degr$\,279 and their uncertainties.

\begin{table}
\caption{Stellar parameters for Sk$-$69$\degr$\,279.}
\label{tab:par}
\begin{center}
\begin{tabular}{lc}
\hline
$\log(L_*/\lsun)$               & 5.54$\pm$0.06    \\
$\log(\dot{M}/\myr)$            & $-$5.26$\pm$0.04 \\
$R_*$ ($\rsun$)                 & 22.7$\pm$2.3   \\
$T_*$ (kK)                      & 29.5$\pm$0.5   \\
$R_{2/3} (\rsun)$               & 35.5$\pm$5.5   \\
$T_{\rm eff}$ (kK)              & $24.0^{+0.5} _{-1.5}$ \\
$v_\infty$ ($\kms$)             & 800$\pm$100 \\
$f_\infty$                      & 0.5 \\
$\beta$                         & 3.0 (fixed) \\
$v\sin i$ ($\kms$)              & 97$\pm$10 \\
$E(B-V)$ (mag)                  & 0.31 (fixed) \\
$M_{\rm bol}$ (mag)             & $-$9.11$\pm$0.15 \\ 
$M_V$ (mag)                     & $-$6.70$\pm$0.15 \\
He (mass fraction)              & 0.50$\pm$0.07 \\
C  (mass fraction)              & 2.3$\pm$1.5$\times10^{-4}$ \\ 
N  (mass fraction)              & 2.8$\pm$0.8$\times10^{-3}$ \\ 
O  (mass fraction)              & 1.2$\pm$0.4$\times10^{-3}$ \\ 
Si (mass fraction)              & 2.4$\pm$0.4$\times10^{-4}$ \\ 
S  (mass fraction)              & 1.0$\pm$0.2$\times10^{-4}$ \\ 
\hline
\end{tabular}
\end{center}
\end{table}

The He/H abundance ratio was estimated by iterative adjustment of this ratio along with other physical parameters of 
the star to reproduce the overall shape of all He and H lines visible in both, UV and optical, spectra. 

The nitrogen abundance was estimated by analysing the behaviour of all nitrogen lines in the optical range. As seen 
from Fig.~\ref{fig:model}, we obtained more or less good fits for the N\,{\sc iii} $\lambda\lambda3999,4004$, 
N\,{\sc iii} $\lambda\lambda4510-47$, N\,{\sc ii} $\lambda\lambda4987,4994$ and 5002--25 lines. On the other hand, 
one can see that in the model spectrum the N\,{\sc iii} $\lambda\lambda4634-40$ lines are in emission, while in the 
observed one they are in absorption. There is also a discrepancy in fitting the N\,{\sc ii} $\lambda\lambda5666-86$ 
lines. These discrepancies may probably be overcome by using a more complicated velocity law for lower parts of the 
wind where the former lines are forming, and by constructing a more expanded atmosphere, because the latter lines appear 
in emission when the temperature decreases to 23 kK.

The O\,{\sc ii} $\lambda\lambda4415,4417$ and O\,{\sc iii} $\lambda5592$ lines were used to estimate the 
oxygen abundance, while the C\,{\sc iv} $\lambda\lambda5801,5812$ and C\,{\sc iii} $\lambda\lambda1175,1247$ lines 
-- to derive the abundance of carbon. We did not use other carbon lines, e.g. C\,{\sc iii} $\lambda\lambda4647,4650,4652$ 
and C\,{\sc iii} $\lambda5696$, to derive the abundance of this element because they are sensitive to a variation of 
$T_{\rm eff}$, $\dot{M}$ and the surface gravity $\log g$, and to the inclusion of other ions, e.g., Fe\,{\sc iv}, 
Fe\,{\sc iv} and S\,{\sc iv}, in calculations (Martins \& Hillier 2012). We also determined the abundances of silicon 
and sulphur using lines of these elements detected in the HRS spectrum. For the abundances of phosphorus and iron-group 
elements we adopted half-solar values because of the low metallicity of the LMC ($Z=Z_\odot /2$).

\section{Spectroscopic observations of the circular shell}
\label{sec:shell}

As noted in Section\,\ref{sec:spe}, the RSS spectrum of the nebula was obtained with the slit oriented south-north 
(PA=0$\degr$) and placed on the star. In the two-dimensional (2D) spectrum of the shell we detected the emission lines of 
H$\gamma$, H$\beta$, H$\alpha$, He\,{\sc i} $\lambda$5876, [N\,{\sc ii}] $\lambda\lambda$6548, 6584 and [S\,{\sc ii}] 
$\lambda\lambda$6717, 6731. All of them are visible on both sides of Sk$-$69$\degr$\,279. The [O\,{\sc iii}] 
$\lambda\lambda$4959, 5007 lines are absent, which indicates a low ionization state in the shell.

\begin{figure}
\begin{center}
\includegraphics[width=8cm,angle=0,clip=]{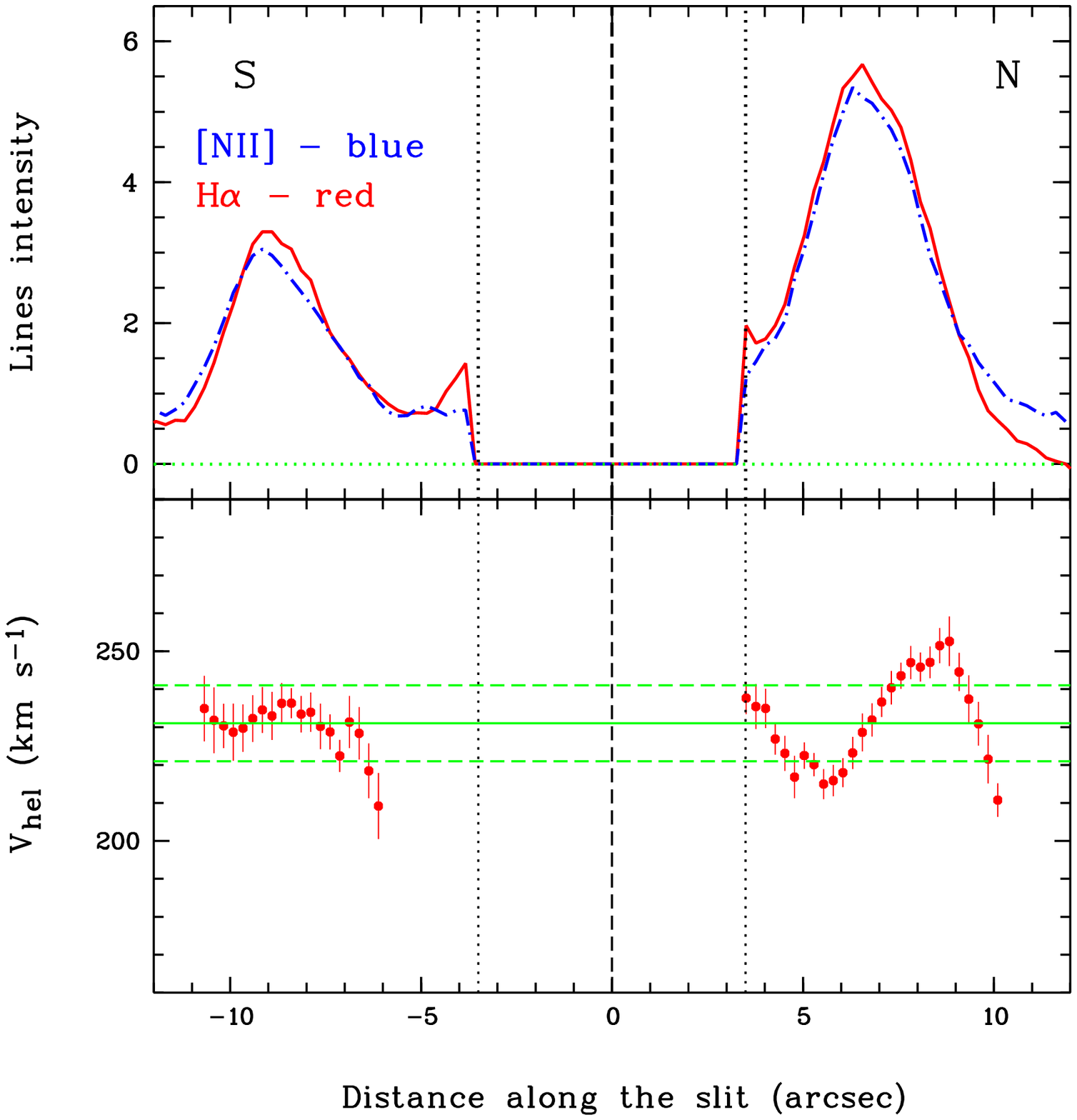}
\end{center}
\caption{Upper panel: Variation along the slit of the H$\alpha$ and [N\,{\sc ii}] 
$\lambda$6584 emission line intensities in the spectrum of the circular shell, 
shown, respectively, with a solid (red) and dashed (blue) lines. Bottom panel: 
H$\alpha$ heliocentric radial velocity distribution along the slit. The (green) 
solid and dashed horizontal lines show, respectively, the systemic velocity of 
the shell of $\approx230 \, \kms$ and its error margins. The dashed vertical line 
in both panels corresponds to the position of Sk$-$69$\degr$\,279, while the dotted 
vertical lines at $\pm$3.5 arcsec from the central line mark the area where the 
emission line intensities and the H$\alpha$ radial velocity were not measured 
because of the effect of the star. S--N direction of the slit is shown.}
\label{fig:ha}
\end{figure}

The upper panel in Fig.\,\ref{fig:ha} plots the H$\alpha$ and [N\,{\sc ii}] 
$\lambda$6584 emission line intensities along the slit. Both lines appear everywhere along 
the slit and their peak intensities show clear correlation with the shell. In the northern 
direction the slit crosses a region of enhanced brightness (identified as knot N in Weiss et al. 
1997), which is manifested in a factor of two higher intensity of the lines.
This panel also shows that the star is offset by $\approx1$ arcsec towards the brighter
part of the shell. The offset and the brightness asymmetry might be caused by either interstellar 
medium density gradient in the north-south direction (cf. Section\,\ref{sec:sk}) or by stellar 
motion to the north (cf. Section\,\ref{sec:dis}), or by both effects. The bottom panel in 
Fig.\,\ref{fig:ha} shows the distribution of the H$\alpha$ heliocentric radial velocity along the slit. 
Using this distribution, we estimated the systemic heliocentric radial velocity of the shell to be 
$\approx230\pm10 \, \kms$, which agrees with that measured by Weis et al. (1997),  
and is offset by $20 \, \kms$ with respect to the background \hii region.

\begin{figure}
\begin{center}
\includegraphics[width=6cm,angle=270,clip=]{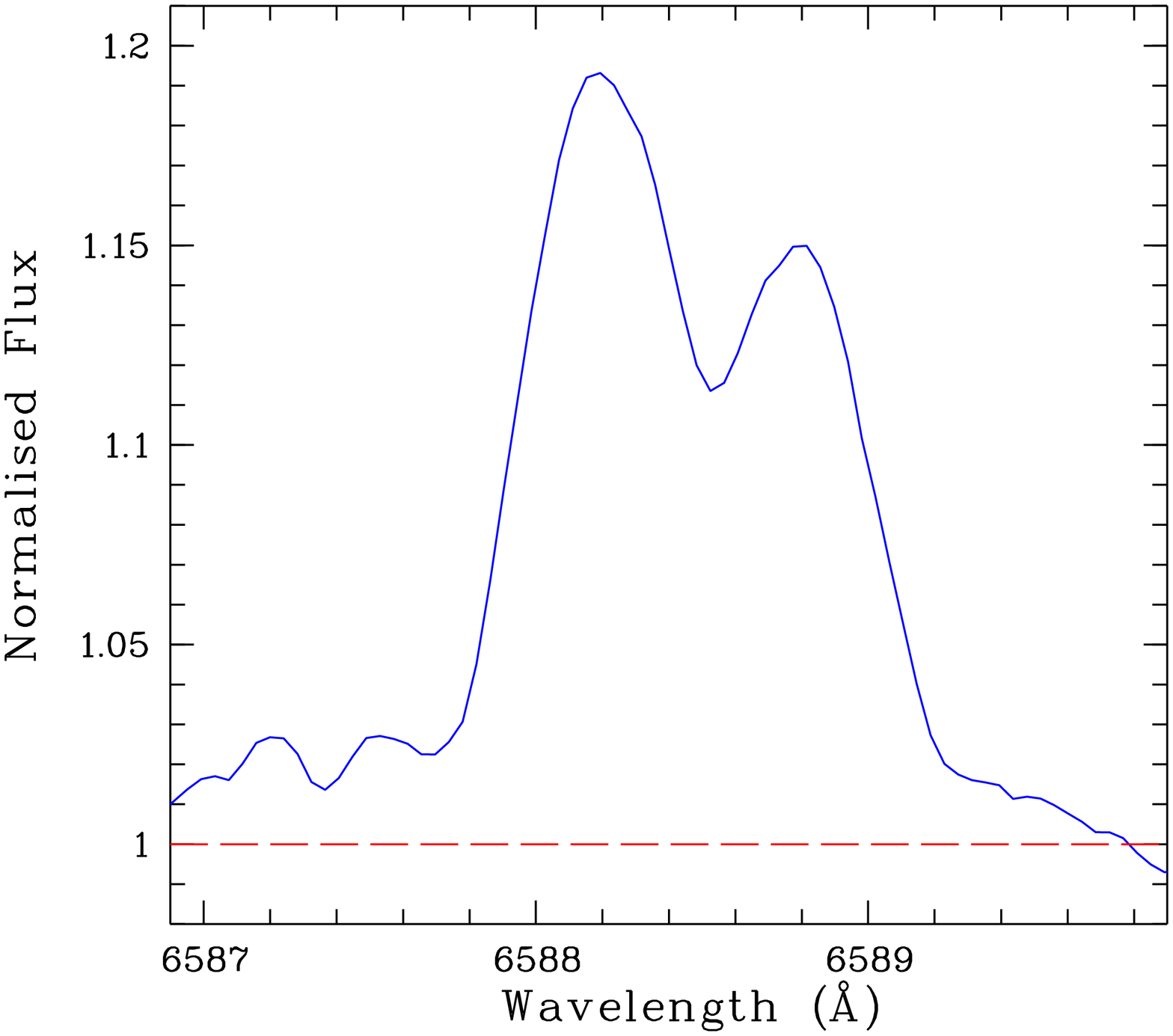}
\end{center}
\caption{Double-peak profile of the [N\,{\sc ii}] $\lambda$6584 line.}
\label{fig:nii}
\end{figure}

The nebular forbidden lines of [N\,{\sc ii}] are also visible in the HRS spectra of Sk$-$69$\degr$\,279, where 
they are resolved in two components (see Fig.\,\ref{fig:nii}), originating in the approaching 
and receding sides of the circumstellar shell. The heliocentric radial velocities of these components are, respectively, 
215.3$\pm$0.5 and 243.7$\pm0.6 \, \kms$, implying the systemic velocity of the shell of $\approx230 \, \kms$, 
which is in a good agreement with the estimate based on the RSS spectrum. 

\begin{figure}
 \includegraphics[angle=-90,width=8.5cm,clip=]{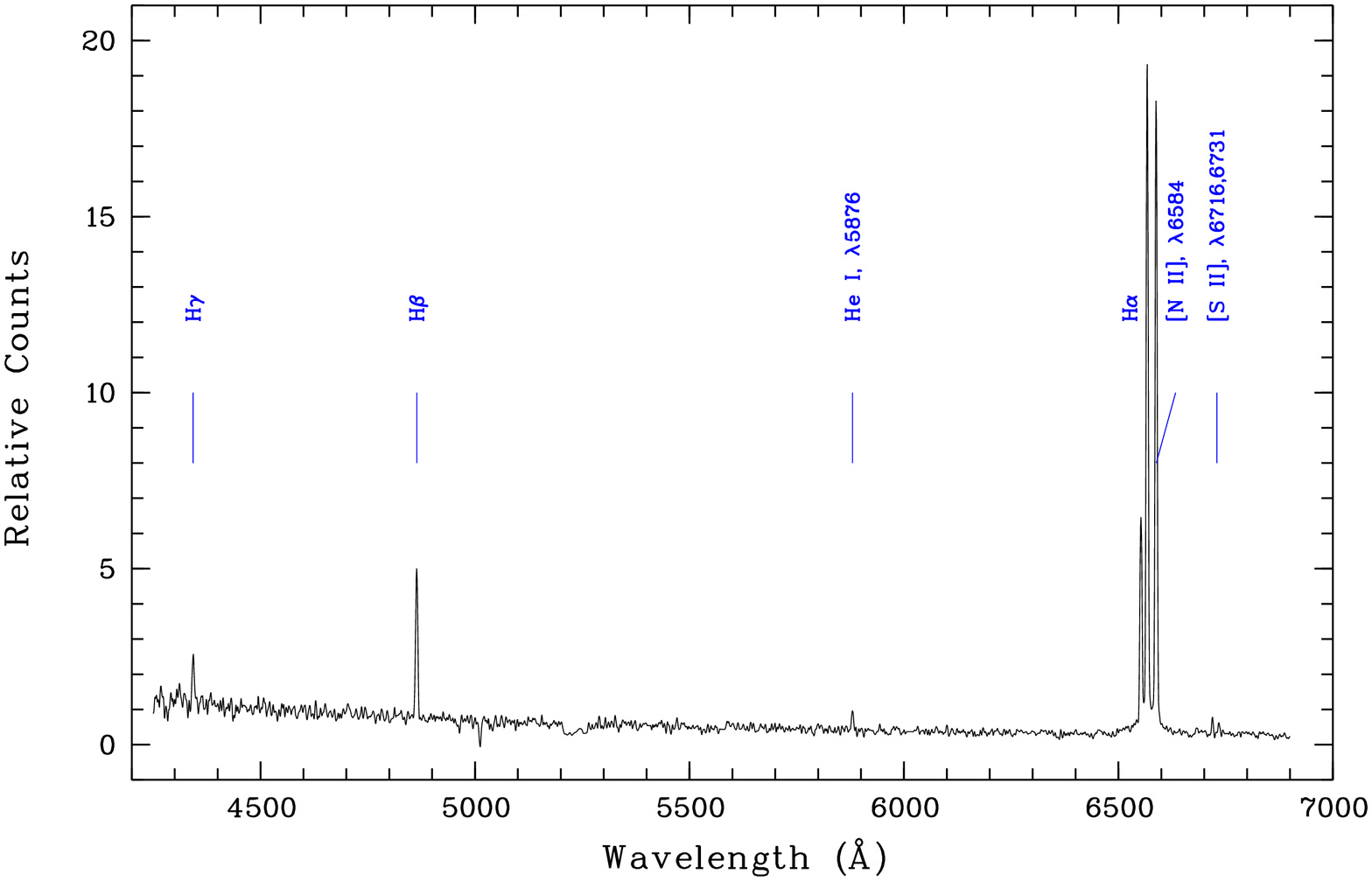}
 \caption{1D reduced spectrum of the circular shell obtained with
  the SALT. All detected emission lines are marked.}
 \label{fig:1D}
\end{figure}

1D RSS spectra of the shell were extracted from a region of radius of 
10 arcsec centred on Sk$-$69$\degr$\,279 with the central $\pm$3.5 arcsec 
excluded. The resulting spectrum is presented in Fig.\,\ref{fig:1D}. Table\,\ref{tab:int}
gives the observed intensities of all detected lines normalized to
H$\beta$, $F(\lambda)/F$(H$\beta$), as well as the reddening-corrected line
intensity ratios, $I(\lambda)/I$(H$\beta$), and the logarithmic
extinction coefficient, $C$(H$\beta$). The latter corresponds to the colour excess of
$E(B-V)$=0.34$\pm$0.02 mag. The lines in Table\,\ref{tab:int} were measured with program
described in Kniazev et al. (2004). We also derived the electron number density using the 
the [S\,{\sc ii}] $\lambda\lambda$6716, 6731 lines, $n_{\rm e}$([S\,{\sc ii}]) and 
added then to Table\,\ref{tab:int}. The obtained density is at the low end of the range of
densities measured for circumstellar nebulae around LBVs (Nota et a. 1995; 
Smith et al. 1998), which is understandable given the large extent
of the shell (cf. Weis et al. 1997). To calculate $C$(H$\beta$) and $n_{\rm e}$([S\,{\sc ii}]),
we assumed that $T_{\rm e}=10^4$ K; for lower values of $T_{\rm e}$ the derived figures 
remain almost the same.

\begin{table}
\centering{ \caption{Line intensities of the circular shell
around Sk$-$69\,279.} \label{tab:int}
\begin{tabular}{lll}
\hline \rule{0pt}{10pt}
$\lambda_{0}$(\AA) Ion                  &
F($\lambda$)/F(H$\beta$)&I($\lambda$)/I(H$\beta$) \\ 
\hline
4340\ H$\gamma$\                        & 0.367$\pm$0.028 & 0.427$\pm$0.033 \\
4861\ H$\beta$\                         & 1.000$\pm$0.038 & 1.000$\pm$0.038 \\
5876\ He\ {\sc i}\                      & 0.134$\pm$0.013 & 0.105$\pm$0.010 \\
6548\ [N\ {\sc ii}]\                    & 1.333$\pm$0.037 & 0.912$\pm$0.028 \\
6563\ H$\alpha$\                        & 4.305$\pm$0.126 & 2.937$\pm$0.094 \\
6584\ [N\ {\sc ii}]\                    & 4.013$\pm$0.109 & 2.727$\pm$0.082 \\
6717\ [S\ {\sc ii}]\                    & 0.089$\pm$0.010 & 0.059$\pm$0.007 \\
6731\ [S\ {\sc ii}]\                    & 0.068$\pm$0.009 & 0.045$\pm$0.006 \\
  & & \\
C(H$\beta$)         & \MC {2}{l}{0.50$\pm$0.04} \\
$E(B-V)$            & \MC {2}{l}{0.34$\pm$0.02 mag} \\
$n_{\rm e}$([Si\,{\sc ii}]) &  \MC {2}{l}{$94.0^{+289.0} _{-84.0} \, {\rm cm}^{-3}$} \\
$v_{\rm hel}$       & \MC {2}{l}{$230\pm10 \, \kms$} \\
\hline
\end{tabular}
 }
\end{table}

Using the reddening-corrected intensities of the [N\,{\sc ii}] and [S\,{\sc ii}] lines, we estimated the nitrogen to 
sulphur abundance ratio, which, according to Benvenuti, D'Odorico \& Peimbert (1973; and references therein) is 
given by:
\begin{eqnarray}
{N({\rm N}^+ )\over N({\rm S}^+ )} =3.61 {I(6584) \over I(6716+6731)} \, . \nonumber 
\end{eqnarray}
This ration is almost independent of $n_{\rm e}$ and $T_{\rm e}$, provided that these parameters are, respectively, 
$\leq 1000 \, {\rm cm}^{-3}$ and $\leq 10^4$\,K. Although our spectrum of the shell did not allow us to estimate $T_{\rm e}$, it is 
reasonable to assume that it is $\leq 10^4$\,K (see, e.g., Nota et al. 1995; Smith et al. 1998). Using the above equation and 
Table\,\ref{tab:int}, one finds 
$N({\rm N}^+ )/N({\rm S}^+ )\approx94.7^{+16.7} _{-13.1}$, where the error margins were derived from the 
uncertainties on the line intensities. Similarly, we estimated the $N({\rm N}^+ )/N({\rm S}^+)$ ratio
in the surrounding \hii region to be 1.42$\pm$0.14, which agrees within the margins of error with the value of 
$2.75^{+2.04} _{-1.17}$ measured for \hii regions in the LMC (Russell \& Dopita 1992). Since the S abundance 
do not change much in the course of stellar evolution, one can conclude that the N abundance in the shell is
elevated by a factor of $\approx34^{+36} _{-17}$ with respect to the LMC value, and that the shell is composed 
mostly of the CNO-processed wind/ejecta material from Sk$-$69$\degr$\,279 (cf. Weis et al. 1997). This conclusion
is supported by the estimate of the surface N abundance of Sk$-$69$\degr$\,279 (see Table\,\ref{tab:par}), which is 
enhanced by about the same factor as in the shell (see next section). Using the LMC abundances of N and S from 
Russell \& Dopita (1992), we derive a N abundance of the shell of $\log({\rm N/H})+12=8.68$. We caution that this 
value could be somewhat overestimated if the shell contains some amount of S$^{++}$.

\section{Discussion}
\label{sec:dis}

\begin{figure*}
 \includegraphics[angle=0,width=11cm,clip=]{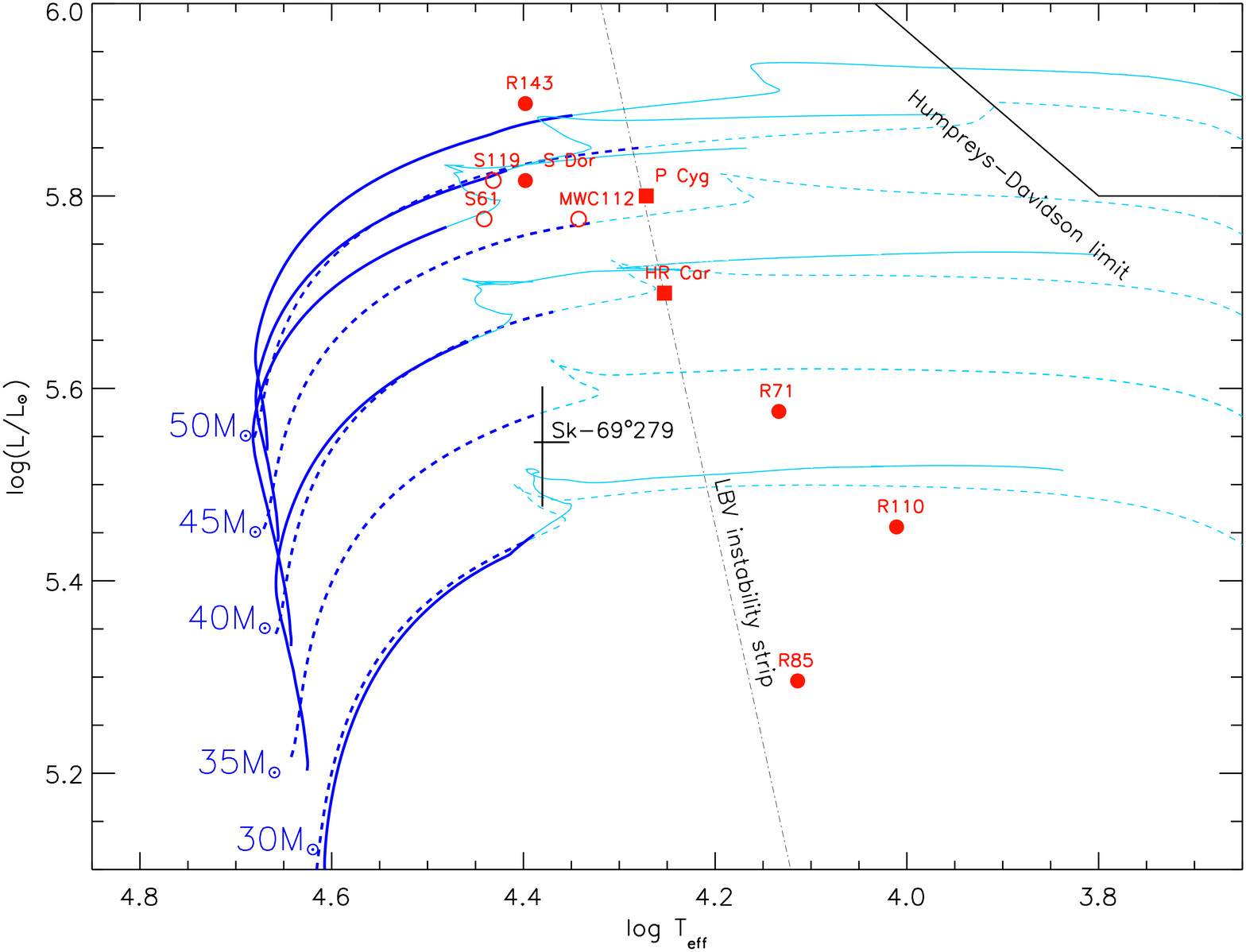}
 \caption{Position of Sk$-$69$\degr$\,279 in the Hertzsprung-Russell diagram. The LBV minimum instability strip 
 is indicated by a dot dashed line (as defined in Groh et al. 2009) and the Humphreys-Davidson limit is shown 
 with a solid (black) line. Positions of two Galactic bona fide LBVs P\,Cygni and HR\,Car (during its visual minimum) 
 are shown with squares; their parameters are, respectively, from Najarro (2001) and Groh et al. (2009). Positions of 
 bona fide and candidate LBVs in the LMC (Humphreys et al. 2016) are shown, respectively, with filled and open circles. 
 The evolutionary tracks for rotating main sequence stars at the LMC metallicity are from Brott et al. (2011). The 
 dashed and solid lines correspond, respectively, to initial rotational velocities of $\approx$110 and $400 \, \kms$.  
 The thick (blue) parts of the tracks correspond to the stage when a star burns the hydrogen in its core.
 }
 \label{fig:HR}
\end{figure*}

From Table\,\ref{tab:par} it follows that the surface of Sk$-$69$\degr$\,279 is highly enriched with nitrogen, 
while the total surface C+N+O abundance agrees within the error margins with the LMC metallicity, implying that there is
CNO-processed material on the stellar surface. Fig.\,\ref{fig:HR} shows the position of Sk$-$69$\degr$279 
in the Hertzsprung-Russell diagram along with evolutionary tracks from Brott et al. (2011) for (single) stars with the LMC 
metallicity and two values of the initial rotational velocity, $v_{\rm init}$, of $\approx$100 and $400 \, \kms$. If 
Sk$-$69$\degr$279 was originally born as a single star, then one can infer that its initial mass was $M_{\rm init}\approx30-35 
\, \msun$ and that this star is either near the end of the main sequence phase or has recently left it. The enhanced N abundances 
in atmospheres of main sequence O and early-B stars was discovered by Lyubimkov (1984) and interpreted as an indicator of 
deep internal mixing of the CNO cycle products. Since the beginning of 2000's the mixing is included in stellar evolution 
models (e.g. Heger \& Langer 2000; 
Meynet \& Maeder 2000) and it is believed that it is caused by the stellar rotation. Whether the rotational mixing is the main 
cause of enhanced surface N abundance in some OB stars is, however, still a subject of debate. Besides the fast stellar rotation, 
the surface N abundance could also be enhanced because of binary interaction processes (e.g. Langer 2012; see also below). 

\begin{table*}
\caption{Comparison of Sk$-$69$\degr$279 with three model stars from
Brott et al. (2011). See the text for details.} 
\label{tab:brott}
\renewcommand{\footnoterule}{}
\begin{center}
\begin{tabular}{lccccc}
\hline 
 & Sk$-$69$\degr$279  & model 1 & model 2 & model 3 \\
\hline
$M_{\rm init} (\msun)$        & --               & 30      & 30           & 35     \\
$v_{\rm init} (\kms)$         & --               & 407     & 439          & 362    \\     
age (Myr)                     & --               & 5.9     & 6.5          & 5.9          \\
$T_{\rm eff}$ (kK)            & 22.5--24.5       & 22.5    & 23.0         & 22.7        \\
$\log(L_*/\lsun)$             & 5.48--5.60       & 5.50    & 5.57         & 5.58    \\
$R_{2/3} (\rsun)$             & 30--41           & 37      & 38           & 40     \\
$\log(\dot{M}/\myr)$          & $-$5.26$\pm$0.04 & $-$5.47 & $-$5.22      & $-$5.43  \\
$v_{\rm surf} (\kms)$         & --               & 233     & 521          & 191   \\  
He (mass fraction)            & 0.5              & 0.3     & 0.4          & 0.3        \\
C (mass fraction) (10$^{-4}$) & 0.8--3.8         & 1.3     & 0.5          & 1.7          \\
N (mass fraction) (10$^{-3}$) & 2.0--3.6         & 1.8     & 2.7          & 1.5          \\ 
O (mass fraction) (10$^{-3}$) & 0.8--1.6         & 1.1     & 0.3          & 1.3          \\
$\Delta$N                     & 20--35           & 18      & 27           & 15     \\
\hline
\end{tabular}
\end{center}
\end{table*}

The models by Brott et al. (2011) show that the higher $v_{\rm init}$ and $M_{\rm init}$ 
of a star the higher its surface He and N abundances. For stars with $M_{\rm init}\approx30-35 \, \msun$ and $v_{\rm 
init}\ga400 \, \kms$ the rotationally induced mixing can enhance these abundances, respectively, by factors of 
$\sim2-3$ and 30 well before the stellar surface will become enriched by fusion products because of the first dredge-up. 

In Table\,\ref{tab:brott}, we compare Sk$-$69$\degr$279 with three model stars from Brott et al. (2011) with the following 
initial masses and rotational velocities: $M_{\rm init}=30 \, \msun$, $v_{\rm init}=407 \, \kms$ (hereafter model\,1), $M_{\rm 
init}=30 \, \msun$, $v_{\rm init}=439 \, \kms$ (model\,2), and $M_{\rm init}=35 \, \msun$, $v_{\rm init}=362 \, \kms$ 
(model\,3). The last row of the table gives factors by which the N abundances of Sk$-$69$\degr$279 and the model 
stars are enhanced with respect to the LMC N abundance of $1.0\times10^{-4}$ (a value adopted in Brott et al. 2011). 
One can see that all three models reproduce fairly good the main parameters of Sk$-$69$\degr$279, such as $T_{\rm 
eff}$, $L_*$ and the surface abundances, and suggest that this star is $\approx6-6.5$ Myr old. 
In models\,1 and 3, the predicted $\dot{M}$ and the He and N abundances are somewhat lower than the 
values derived for Sk$-$69$\degr$279. These quantities, however, increase with the increase of $v_{\rm init}$ and in model\,2
they better match the corresponding parameters of Sk$-$69$\degr$279 (the C and O abundances predicted by this model, however,
are lower than in Sk$-$69$\degr$279). Interestingly, when the star transitions from core-H to core-He burning and contracts, 
model\,2 shows a factor of 10 increase in $\dot{M}$ on a time scale of $\sim1\,000$ yr\footnote{This increase in $\dot{M}$
is related to the bi-stability jump (Vink, de Koter \& Lamers 2000) and spinning up of the surface layers caused by the overall 
contraction of the star.}, while other stellar parameters remain almost intact and comparable to those of Sk$-$69$\degr$279, except 
of the surface rotational velocity, $v_{\rm surf}$, which slows down by $\approx30$ per cent on the same time scale. For even higher 
$v_{\rm init}$, the mass fractions of He and N on the surface of a $30 \, \msun$ star increase, respectively, to $\approx$0.8 and 
$3\times10^{-3}$. Such fast-spinning stars, however, are much hotter than Sk$-$69$\degr$279 and remain always in the blue part of the 
Hertzsprung-Russell diagram.

Table\,\ref{tab:brott} also shows that $v_{\rm surf}$ of all three model stars is higher than $v\sin i$ measured for
Sk$-$69$\degr$279. Although we cannot exclude the possibility that Sk$-$69$\degr$279 is oriented nearly pole-on and therefore 
its rotational velocity is actually higher, it is also possible that this (presumably initially fast-spinning) star has reached 
the limit of hydrostatic stability ($\Omega$-limit; Langer 1997, 1998) and ejected its outer layers either instantly or during 
a brief episode of enhanced mass loss. If so, this might be responsible for the formation of the circumstellar shell and 
slowing down the stellar surface and its enrichment in helium and nitrogen. The origin of the shell around 
Sk$-$69$\degr$279 could also be due to enhanced mass loss caused by the bi-stability jump, while the presumably high 
initial rotational velocity of this star may be significantly reduced because of the bi-stability braking (Vink et al. 2010).

While the major parameters of Sk$-$69$\degr$279 could be explained fairly well in the framework of rapidly rotating 
single star models, it is also possible that this star is the product of binary evolution and that its
enhanced He and N abundances are because of accretion of CNO-processed material from the binary companion star or due to merger 
with this star. Both these processes could also be responsible for spinning up and rotational mixing of Sk$-$69$\degr$279, 
and thereby for enhanced mass-loss and additional enrichment of the stellar surface in He and N. According to Sana et al. 
(2012, 2013), the majority of massive stars are members of binary systems and the evolution of many of them is strongly affected 
by interaction with a companion star already during the main sequence stage (see also de Mink et al. 2014).

The above considerations support the proposal by Lamers et al. (2001) that fast-rotating stars could produce LBV-like nebulae 
while they are at or near the end of the main sequence phase. The same conclusion was also drawn from studies of 
the blue supergiants Sher\,25 (B1.5 Iab; Smartt et al. 2002; Hendry et al. 2008), TYC\,3159-6-1 (O9.5--O9.7\,Ib; Gvaramadze et al. 
2014a), [GKF2010]\,MN18 (B1\,Ia; Gvaramadze et al. 2015) and HD\,168625 (B6\,Iap; Mahy et al. 2016), all of which are associated 
with compact circumstellar nebulae. The bipolar morphology of these nebulae strongly suggest that their central stars were 
nearly critical rotators in the past (the projected rotational velocities of these stars are similar to that of Sk$-$69$\degr$279), 
while the CNO abundances of the stars indicate that they are either near the end of the main sequence or just evolved off it. 

The ejection of the outer layers of fast-spinning stars can not only result in the origin of circumstellar nebulae and 
slowing down these stars but, presumably, may also be accompanied by the LBV-like variability. The bloated atmosphere of 
Sk$-$69$\degr$279 and its moderate wind velocity may, therefore, be residual of the LBV activity in the recent past.

To conclude, we note that the $20 \, \kms$ difference between the radial velocities of the circumstellar 
shell and the local ISM hints at the possibility that Sk$-$69$\degr$279 is a 
runaway star (cf. Danforth \& Chu 2001). This possibility is supported by the 
isolated location of this star from known star clusters (cf. Gvaramadze et al. 2012b).
If the brightness asymmetry of the circumstellar shell around Sk$-$69$\degr$279 is caused by 
motion of this star to the north, then its possible birth place is the open star cluster
[M87]\,OB\,6 (Melnick 1987), located at about 5 arcmin (or $\approx$70 pc in projection) to the south.
The age of this cluster of $15^{+15} _{-10}$ Myr (Glatt, Grebel \& Koch 2010) agrees within the margins of error 
with the age of Sk$-$69$\degr$279 of $\approx$6--6.5 Myr.

\section{Acknowledgements}

This work is based on observations obtained with the Southern African Large Telescope (SALT), programmes 
\mbox{2011-3-RSA\_OTH-002}, \mbox{2016-1-SCI-012} and \mbox{2017-1-SCI-006}, and supported by the Russian Foundation 
for Basic Research grant 16-02-00148. AYK and OVM acknowledge support from, respectively, the National Research 
Foundation (NRF) of South Africa and the project RVO:67985815 in the Czech Republic. We are grateful to 
F.\,Martins for providing us the {\sc cmfgen} model of an O9\,I star.
Some of the data presented in this paper were obtained from the Mikulski Archive for Space Telescopes 
(MAST), the Digital Access to a Sky Century @ Harvard (DASCH), and the OMC Archive at CAB (INTA-CSIC), 
pre-processed by ISDC. STScI is operated by the Association of Universities for Research in Astronomy, Inc., 
under NASA contract NAS5-26555. Support for MAST for non-HST data is provided by the NASA Office of Space 
Science via grant NNX09AF08G and by other grants and contracts. The DASCH project at Harvard is grateful for 
partial support from NSF grants AST-0407380, AST-0909073, and AST-1313370. This work has made use of the 
NASA/IPAC Infrared Science Archive, which is operated by the Jet Propulsion Laboratory, California Institute 
of Technology, under contract with the National Aeronautics and Space Administration, the SIMBAD data base 
and the VizieR catalogue access tool, both operated at CDS, Strasbourg, France.


\begin{thebibliography}{}
%
\bibitem{} Alfonso-Garzon J., Domingo A., Mas-Hesse J. M., 2010, Proc. Sci., First
catalogue of variable sources observed by OMC onboard INTEGRAL.
SISSA, Treiste, PoS(INTEGRAL 2010)069
\bibitem{} Barnes S. I. et al., 2008, in McLean I. S., Casali M. M., eds, Proc. SPIE
Conf. Ser. Vol. 7014, Ground-based and Airborne Instrumentation for
Astronomy II. SPIE, Bellingham, p. 70140K
\bibitem{} Benvenuti P., D'Odorico S., Peimbert M., 1973, A\&A, 28, 447
\bibitem{} Berdnikov L., Vozyakova O. V., Kniazev A. Yu., Kravtsov V. V., Dambis
A. K., Zhuiko S. V., 2012, Astron. Rep., 56, 290
\bibitem{} Bohannan B., 1997, in Nota A., Lamers H. J. G. L. M., eds, ASP Conf. Ser.
Vol. 120, Luminous Blue Variables:Massive Stars in Transition. Astron.
Soc. Pac., San Francisco, p. 120
\bibitem{} Bohannan B., Epps H. W., 1974, A\&AS, 18, 47
\bibitem{} Bonanos A. Z. et al., 2009, AJ, 138, 1003
\bibitem{} Bramall D. G. et al., 2010, in McLean I. S., Ramsay S. K., Takami H., eds,
Proc. SPIE Conf. Ser. Vol. 7735, Ground-based and Airborne Instrumentation
for Astronomy III. SPIE, Bellingham, p. 77354F
\bibitem{} Bramall D. G. et al., 2012, in McLean I. S., Ramsay S. K., Takami H., eds,
Proc. SPIE Conf. Ser. Vol. 8446, Ground-based and Airborne Instrumentation
for Astronomy IV. SPIE, Bellingham, p. 84460A
\bibitem{} Brott I. et al., 2011, A\&A, 530, A115
\bibitem{} Buckley D. A. H., Swart G. P., Meiring J. G., 2006, in Stepp L. M., ed.,
Proc. SPIEConf. Ser.Vol. 6267, Ground-based and AirborneTelescopes.
SPIE, Bellingham, p. 62670Z
\bibitem{} Burgh E. B., Nordsieck K. H., Kobulnicky H. A., Williams T. B.,
O'Donoghue D., Smith M. P., Percival J. W., 2003, in Iye M., Moorwood
A. F. M., eds, Proc. SPIE Conf. Ser. Vol. 4841, Instrument Design
and Performance for Optical/Infrared Ground-based Telescopes. SPIE,
Bellingham, p. 1463
\bibitem{} Clark J. S., Larionov V. M., Arkharov A., 2005, A\&A, 435, 239
\bibitem{} Clark J. S., Egan M. P., Crowther P. A., Mizuno D. R., Larionov V. M.,
Arkharov A., 2003, A\&A, 412, 185
\bibitem{} Conti P. S., Alschuler W. R., 1971, ApJ, 170, 325
\bibitem{} Conti P. S., Garmany C. D., Massey P., 1986, AJ, 92, 48
\bibitem{} Crause L. A. et al., 2014, in Ramsay S. K., McLean I. S., Takami H., eds,
Proc. SPIE Conf. Ser. Vol. 9147, Ground-based and Airborne Instrumentation
for Astronomy V. SPIE, Bellingham, p. 91476T
\bibitem{} Crawford S. M. et al., 2010, in Silva D. R., Peck A. B., Soifer B. T., eds,
Proc. SPIE Conf. Ser. Vol. 7737, Observatory Operations: Strategies,
Processes, and Systems III. SPIE, Bellingham, p. 773725
\bibitem{} Crowther P. A., Lennon D. J., Walborn N. R., 2006, A\&A, 446, 279
\bibitem{} Cutri R.M. et al., 2003, VizieR Online Data Catalog, 2246, 0
\bibitem{} Danforth C.W., Chu Y.-H., 2001, ApJ, 552, L155
\bibitem{} de Mink S. E., Sana H., Langer N., Izzard R. G., Schneider F. R. N., 2014, ApJ, 782, 7
\bibitem{} Evans C. J., Lennon D. J., Walborn N. R., Trundle C., Rix S. A., 2004, PASP, 116, 909
\bibitem{} Fazio G. G. et al., 2004, ApJS, 154, 10
\bibitem{} Gibson B. K., 2000, Mem. Soc. Astron. Ital., 71, 693
\bibitem{} Glatt K., Grebel E. K., Koch A., 2010, A\&A, 517, A50
\bibitem{} Grindlay J., Tang S., Simcoe R., Laycock S., Los E., Mink D., Doane A.,
Champine G., 2009, in Osborn W., Robbins L., eds, ASP Conf. Ser. Vol.
410, Preserving Astronomy's Photographic Legacy: Current State and
the Future of North American Astronomical Plates. Astron. Soc. Pac.,
San Francisco, p. 101
\bibitem{} Groh J. H., Hillier D. J., Damineli A., Whitelock P. A., Marang F., Rossi C., 2009, ApJ, 698, 1698
\bibitem{} Groh J. H. et al., 2009, ApJ, 705, L25
\bibitem{} Gvaramadze V. V., Kniazev A. Y., 2017, in Miroshnichenko A.S., Zharikov S.V., Korcakova
D., Wolf M., eds., ASP Conf. Ser. Vol. 508, The B[e] Phenomenon. Forty Years of Studies. Astron. 
Soc. Pac., San Francisco, p. 207 
\bibitem{} Gvaramadze V. V., Kniazev A. Y., Fabrika S., 2010, MNRAS, 405, 1047
\bibitem{} Gvaramadze V. V., Kroupa P., Pflamm-Altenburg J., 2010, A\&A, 519, A33
\bibitem{} Gvaramadze V. V., Pflamm-Altenburg J., Kroupa P., 2011, A\&A, 525, A17
\bibitem{} Gvaramadze V. V., Weidner C., Kroupa P., Pflamm-Altenburg J., 2012b, MNRAS, 424, 3037
\bibitem{} Gvaramadze V. V., Miroshnichenko A. S., Castro N., Langer N., Zharikov S. V., 2014a, MNRAS, 437, 2761
\bibitem{} Gvaramadze V. V. et al., 2012a, MNRAS, 421, 3325
\bibitem{} Gvaramadze V. V. et al., 2014b, MNRAS, 442, 929
\bibitem{} Gvaramadze V. V. et al., 2015, MNRAS, 454, 219
\bibitem{} Heger A., Langer N., 2000, ApJ, 544, 1016
\bibitem{} Hendry M. A., Smartt S. J., Skillman E. D., Evans C. J., Trundle C., Lennon D. J., Crowther P. A., Hunter I., 
2008, MNRAS, 388, 1127
\bibitem{} Hillier D. J., Miller D. L., 1998, ApJ, 496, 407
\bibitem{} Hillier D. J., Miller D. L., 1999, ApJ, 519, 354
\bibitem{} Hillier D. J., Lanz T., Heap S. R., Hubeny I., Smith L. J., Evans C. J., Lennon D. J., Bouret J.-C., 2003, ApJ, 588, 1039
\bibitem{} Howarth I. D., 1983, MNRAS, 203, 301
\bibitem{} Hubeny I., Lanz T., 1995, ApJ, 439, 875
\bibitem{} Humphreys R. M., Weis K., Davidson K., Gordon M. S., 2016, ApJ, 825, 64
\bibitem{} Isserstedt J., 1975, A\&AS, 19, 259
\bibitem{} Kerton C. R., Ballantyne D. R., Martin P. G., 1999, AJ, 117, 2485
\bibitem{} Kniazev A. Y., Gvaramadze V. V., 2015, in Proceedings of the SALT Science Conference 2015 (SSC2015). 
Stellenbosch Institute of Advanced Study, South Africa, id. 49
\bibitem{} Kniazev A. Y., Gvaramadze V. V., Berdnikov L. N., 2015, MNRAS, 449, L60
\bibitem{} Kniazev A. Y., Gvaramadze V. V., Berdnikov L. N., 2016, MNRAS, 459, 3068
\bibitem{} Kniazev A. Y., Pustilnik S. A., Grebel E. K., Lee H., Pramskij A. G., 2004, ApJS, 153, 429
\bibitem{} Kniazev A. Y., Grebel E. K., Pustilnik S. A., Pramskij A. G., Zucker D. B., 2005, AJ, 130, 1558
\bibitem{} Kniazev A. Y. et al., 2008, MNRAS, 388, 1667
\bibitem{} Kobulnicky H. A., Nordsieck K. H., Burgh E. B., Smith M. P., Percival
J. W., Williams T. B., O'Donoghue D., 2003, in Iye M., Moorwood
A. F. M., eds, Proc. SPIE Conf. Ser. Vol. 4841, Instrument Design
and Performance for Optical/Infrared Ground-based Telescopes. SPIE,
Bellingham, p. 1634
\bibitem{} Kudritzki R. P., Puls J., Lennon D. J., Venn K. A., Reetz J., Najarro F., McCarthy J. K., Herrero A., 1999, A\&A, 350, 970
\bibitem{} Lamers H. J. G. L. M., Pauldrach A. W. A., 1991, A\&A, 244, L5
\bibitem{} Lamers H. J. G. L. M., Snow T. P., Lindholm D. M., 1995, ApJ, 455, 269
\bibitem{} Lamers H. J. G. L. M., Cassinelly J. P., 1999, Introduction to stellar winds, Cambridge University Press
\bibitem{} Lamers H. J. G. L. M., Nota A., Panagia N., Smith L. J., Langer N., 2001, ApJ, 551, 764
\bibitem{} Langer N., 1997, in Nota A., Lamers H., eds, ASP Conf. Ser. Vol. 120, Luminous Blue Variables: Massive Stars in 
Transition. Astron. Soc. Pac., San Francisco, p. 83
\bibitem{} Langer N., 1998, A\&A, 329, 551
\bibitem{} Langer N., 2012, ARA\&A, 50, 107
\bibitem{} Lanz T., Hubeny I., 2003, ApJS, 146, 417
\bibitem{} Lyubimkov L. S., 1984, Astrophysics, 20, 255
\bibitem{} Mahy L., Hutsem\'ekers D., Royer P., Waelkens C., 2016, A\&A, 594, A94
\bibitem{} Markova N., Puls J., Repolust T., Markov H., 2004, A\&A, 413, 693
\bibitem{} Marston A. P., 1995, AJ, 109, 1839
\bibitem{} Martins F., Plez B., 2006, A\&A, 457, 637
\bibitem{} Martins F., Hillier D. J., 2012, A\&A, 545, A95
\bibitem{} Martins F., Schaerer D., Hillier D. J., 2005, A\&A, 436, 1049
\bibitem{} McLean B. J., Greene G. R., Lattanzi M. G., Pirenne B., 2000, in Manset N., Veillet C., Crabtree D., 
eds, ASP Conf. Ser. Vol. 216, Astronomical Data Analysis Software and
Systems IX. Astron. Soc. Pac., San Francisco, p. 145
\bibitem{} Meixner M. et al., 2006, AJ, 132, 2268
\bibitem{} Melnick J., 1987, in Khachikian E. E., Fricke K. J., Melnick J., eds, IAU Symp. 121, Observational Evidence of 
Activity in Galaxies. Kluwer Academic Publishers, Dordrecht, p. 545
\bibitem{} Meynet G., Maeder A., 2000, A\&A, 361, 101
\bibitem{} Mokiem M. R. et al., 2007, A\&A, 473, 603
\bibitem{} Morrell N. I., Walborn N. R., Fitzpatrick E. L., 1991, PASP, 103, 341
\bibitem{} Najarro F., 2001, in de Groot M., Sterken C., eds, ASP Conf. Ser. Vol. 233, P Cygni 2000: 400 Years 
of Progress. Astron. Soc. Pac., San Francisco, p. 133
\bibitem{} Najarro F., Hanson M. M., Puls J., 2011, A\&A, 535, A32
\bibitem{} Nandy K., Thompson G. I., Morgan D. H., Houziaux L., 1984, MNRAS, 210, 131
\bibitem{} Nota A., Livio M., Clampin M., Schulte-Ladbeck R., 1995, ApJ, 448, 788
\bibitem{} O'Donoghue D. et al., 2006, MNRAS, 372, 151
\bibitem{} Pauldrach A. W. A., Puls J., 1990. A\&A, 237. 409
\bibitem{} Penny L. R., Gies D. R., 2009, ApJ, 700, 844
\bibitem{} Pilbratt G. L. et al., 2010, A\&A, 518, L1
\bibitem{} Pojma\'nski, G., 2002, Acta Astron., 52, 397
\bibitem{} Puls J., Markova N., Scuderi S., Stanghellini C., Taranova O. G., Burnley A. W., Howarth I. D., 2006, A\&A, 454, 625
\bibitem{} Puls J. et al., 1996, A\&A, 305, 171
\bibitem{} Rieke G. H. et al., 2004, ApJS, 154, 25
\bibitem{} Rousseau J., Martin N., Pr\'{e}vot L., Rebeirot E., Robin A., Brunet J. P., 1978, A\&AS, 31, 243
\bibitem{} Russell S. C., Dopita M. A., 1992, ApJ, 384, 508
\bibitem{} Sana H. et al., 2012, Science, 337, 444
\bibitem{} Sana H. et al., 2013, A\&A, 550, A107
\bibitem{} Sanduleak N., 1970, Contribution, La Serena: Cerro Tololo, 89
\bibitem{} Smartt S. J., Lennon D. J., Kudritzki R. P., Rosales F., Ryans R. S. I., Wright N., 2002, A\&A, 391, 979
\bibitem{} Smith N., Vink J. S., de Koter A., 2004, ApJ, 615, 475
\bibitem{} Smith L. J., Nota A., Pasquali A., Leitherer C., Clampin M., Crowther P. A., 1998, ApJ, 503, 278
\bibitem{} Smith Neubig M. M., Bruhweiler F. C., 1999, AJ, 117, 2856 
\bibitem{} Sota A., Ma\'{i}z-Apell\'{a}niz J., Walborn N.R., Alfaro E.J., Barb\'{a} R.H., Morrell N.I., Gamen R.C., 
Arias J.I., 2011, ApJS, 139, 24
\bibitem{} Sota A., Ma\'{i}z-Apell\'{a}niz J., Morrell N. I., Barb\'a R. H., Walborn N. R.,
Gamen R. C., Arias J. I., Alfaro E. J., 2014, ApJS, 211, 10
\bibitem{} Steele I. A., Negueruela I., Clark J. S., 1999, A\&AS, 137, 147
\bibitem{} \v{S}urlan B., Hamann W.-R., Kub\'at J., Oskinova L. M., Feldmeier A., 2012, A\&A, 541, A37
\bibitem{} \v{S}urlan B., Hamann W.-R., Aret A., Kub\'at J., Oskinova L. M., Torres A. F., 2013, A\&A, 559, A130
\bibitem{} Taylor W. D., Evans C. J., Sim\'on-D\'iaz S., Sana H., Langer N., Smith N., Smartt S. J., 2014, MNRAS, 442, 1483
\bibitem{} Van Genderen A. M., 2001, A\&A, 366, 508
\bibitem{} Vink J. S., de Koter A., Lamers H. J. G. L. M., 1999, A\&A, 350, 181
\bibitem{} Vink J. S., de Koter A., Lamers H. J. G. L. M., 2000, A\&A, 362, 711
\bibitem{} Vink J. S., Brott I., Gr\"afener G., Langer N., de Koter A., Lennon D. J., 2010, A\&A, 512, L7
\bibitem{} Wachter S., Mauerhan J.C., van Dyk S.D., Hoard D.W., Kafka S., Morris P.W., 2010, AJ, 139, 2330
\bibitem{} Weis K., Bomans D. J., Chu Y.-H., Joner M. D., Smith R. C., 1995, in Pena M., Kurtz S., eds, Rev. Mex. Astron. Astrofis.
Ser. Conf., 3, 237
\bibitem{} Weis K., Chu Y.-H., Duschl W. J., Bomans D. J., 1997, A\&A, 325, 1157
\bibitem{} Zaritsky D., Harris J., Thompson I. B., Grebel E. K., 2004, AJ, 128, 1606
\end{thebibliography}
\end{document}